\documentstyle[onecolumn]{mn}

\newcounter{parentequation}
\def\beglet{
  \addtocounter{equation}{1}%
  \setcounter{parentequation}{\value{equation}}%
  \setcounter{equation}{0}%
  \def\theequation{\arabic{parentequation}\alph{equation}}%
  \ignorespaces
}
\def\endlet{
  \setcounter{equation}{\value{parentequation}}%
  \def\theequation{\arabic{equation}}%
}
\def\pmb#1{\setbox0=\hbox{#1}%
    \kern-.025em\copy\kern-\wd0
    \kern.05em\copy\kern-\wd0
    \kern-.025em\raise.0433em\box0}

\def\ltsima{$\; \buildrel < \over \sim \;$}
\def\gtsima{$\; \buildrel > \over \sim \;$}
\def\simlt{\lower.5ex\hbox{\ltsima}}
\def\simgt{\lower.5ex\hbox{\gtsima}}

\def\etal{{\it et al.}\rm}
\def\etals{{\it et al. }\rm}
\def\mk2{\mu {\rm K}^2}
\def\planck{\it Planck\rm}
\def\plancks{\it Planck \rm}

\begin{document}

\title[Polarisation Destriping Errors]
{Effects of Destriping Errors on CMB Polarisation Power Spectra and
and Pixel Noise Covariances}

 \author[G. Efstathiou]{G. Efstathiou\\
Institute of Astronomy, Madingley Road, Cambridge, CB3 OHA.}

\maketitle

\begin{abstract}
Low frequency detector noise in CMB experiments must be corrected to
produce faithful maps of the temperature and polarization
anisotropies. For a \planck-type experiment the low frequency noise
corrections lead to residual stripes in the maps. Here I show that for
a ring torus and idealised detector geometry it is possible to
calculate analytically the effects of destriping errors on the
temperature and polarization power spectra. It is also possible to
compute the pixel-pixel noise covariances for maps of arbitrary
resolution. The analytic model is compared to numerical simulations
using a realistic detector and scanning geometries. We show that
\plancks polarization maps at $143$GHz should be signal dominated on
large scales. Destriping errors are the dominant source of noise for
the temperature and polarization power spectra at multipoles $\ell
\simlt 10$.  A fast Monte-Carlo method for characterising noise,
including destriping errors, is described that can be applied to
\planck. This Monte-Carlo method can be used to quantify pixel-pixel
noise covariances and to remove noise biases in power spectrum estimates.

\vskip 0.1 truein


\vskip 0.3 truein

\end{abstract}

\section{Introduction}

This is the fourth in a series of papers (Efstathiou 2004, 2005, 2006;
henceforth E04, E05, E06 respectively) outlining a methodology for
analysing power spectra for a \planck-type cosmic microwave background
(CMB) experiment\footnote{For an up-to-date account of the \plancks
satellite and its science case see {\it `The Scientific Programme of
\planck'} by the \plancks Consortia (2005) hereafter
SPP05. }. The first of these papers discussed fast hybrid methods for
estimating temperature power spectra. The hybrid method combines a
maximum likelihood estimator at low multipoles, computed from low
resolution maps, with a pseudo-$C_\ell$ (PCL) estimator at high
multipoles computed using fast spherical transforms applied to high
resolution maps (see Hivon \etals 2002 and references therein). A near
optimal hybrid estimator can be constructed if the noise properties of
the input maps obey certain properties, {\it e.g.} if the noise is
accurately white at high multipoles and if correlated noise can be
accounted for in the pixel-pixel noise covariance matrices for the low
resolution maps used in the maximum likelihood estimate. If this is
the case, accurate analytic expressions for the PCL covariance
matrices can be derived for realistic sky cuts. The hybrid estimator
concept was generalised to polarization in E06 and analytic formulae
for PCL polarization power spectra for noise-free and noisy data were
derived. (See also the related paper by Challinor and Chon,
2005). There is a large literature on power spectrum estimation
methods and we will not attempt to give a comprehensive bibliography
here. A fairly complete set of references to previous work can be
found in the papers of hybrid power spectrum estimators (E04, E06).

CMB temperature map reconstruction errors for a \planck-type were
considered in E05, on the assumption that maps would be constructed
using simple `destriping' techniques (see {\it e.g.}  Burigana \etals
1997; Delabrouille 1998; Maino \etals 1999; Revenu \etals 2000;
Keih\"anen \etals 2004, 2005). Destriping algorithms are particularly
suited to a \planck-type scanning strategy in which the sky is scanned
many times on rings.  In E05, it was shown that for a ring torus
scanning geometry (and provided the instrument noise satisfies certain
reasonable properties) it is possible to compute accurately the
effects of residual striping errors on the temperature power spectrum.
For realistic detector noise, the resulting error power spectrum is
indeed accurately white at high multipoles, but at low multipoles the
striping errors introduce a bias in the temperature power spectrum
that has a distinctive `sawtooth' type pattern with an envelope that
varies approximately as $1/\ell$. For realistic \planck-type scanning
geometries, the ring torus model provides a good (but not exact)
description of the error power spectrum determined from simulations
of the time-ordered data (TOD). E05 also argued that for a slowly
precessing scanning strategy, and detector noise properties such as
envisaged for \planck, a simple destriping map making algorithm will
be close to optimal. The destriping errors are dominated by
`irreducable' white noise errors in the ring overlaps and so there is
little to be gained by implementing computationally demanding least
squares map making methods ({\it e.g.}  Wright \etals
1996; Tegmark 1997a,b; Borrill \etals 2001; Natoli \etals 2001; Dor\'e
\etals 2001). This conjecture has been verified by Ashdown \etals
(2006, hereafter A06), who compared a number of map making codes
applied to simulated \plancks data (specifically, simulations of four
217 GHz detectors). The temperature power spectrum residuals from the
different codes are almost identical, except for small differences at
low multipoles. A simple destriping code (for example, the Springtide
code described by A06) fitting one offset (or `baseline') per ring
pointing produces a map that is very close to optimal. Depending on
the value of the detector noise `knee frequency', the accuracy of a
destriping algorithm can be improved further by increasing the number
of baselines and by applying more sophisticated destriping algorithms that
include information on the correlations between baseline offsets
(Keih\"anen, Kurki-Suonio and Poutanen, 2005). However,  the simulations of
A06 show that for the knee frequencies expected for \plancks 
the improvements in the maps are minor as the number of baselines is
increased above one per scanning ring.

In this paper we extend the analysis of E05 to polarization.  As we
will show in Section 3, for a ring torus geometry it is possible to
construct  an analytic model for temperature and polarization
destriping errors. Formulae are presented in Section 3 for the
contribution of residual destriping errors to the temperature and
polarization power spectra and also for the pixel-pixel noise
covariances in maps of arbitrary resolution.

Section 4 describes numerical simulations of a \planck-like focal
plane geometry and slow precessing scanning strategy. Specifically, we
use the geometry of the four polarization sensitive bolometer pairs at
$143$ GHz (see SPP05). However, to make large numbers (thousands) of
simulations feasible, we have assumed $60^\prime$ FWHM beams instead of
the $7.1^\prime$ beams of \planck, and that the scans consist of
$1080$ ring pointings instead of the
$8640$ ring pointings of \planck. The detector white noise
assumed for these simulations was adjusted so that the noise
properties of the maps match those expected from \plancks at equivalent
spatial resolution. These simulations therefore give a good impression
of the actual noise properties of \plancks $143$GHz maps at low
multipoles.  With these simulations it is possible to compute the
pixel-pixel noise covariance matrix for degraded resolution maps and
to isolate the component arising from destriping errors. These pixel
noise covariance matrices can be compared with the ring torus
approximation derived in Section 3. We then develop an extremely fast
and accurate Monte-Carlo method of estimating the pixel-pixel noise
covariance matrices based on the statistical properties of destriping
baseline offsets. This method offers a reliable way of quantifying the
pixel-pixel covariance matrices for temperature and polarization maps
of arbitrary resolution and for realistic noise and scanning
geometries. These pixel-pixel noise covariance estimates can then be
used as inputs to maximum likelihood methods of power-spectrum
estimation (E04, E06, Slosar and Seljak 2004) or map-based Gibbs
sampling methods (see {\it e.g.} Larson \etals 2006).

\section{Overview of Polarisation Map Making with Destriping}

We follow the notation of Keih\"anen \etals (2005, hereafter KKP05) and
write the TOD stream as 
\begin{equation}
{\bf y} = {\bf P}{\bf m} + {\bf n^\prime},  \label{TOD1}
\end{equation}
where ${\bf m}$ is the pixelised CMB temperature map, ${\bf P}$ is a
pointing matrix, and ${\bf n^\prime}$ is the detector
noise. (Polarized maps will be dealt with below, but to keep the
equations simple we will assume for the moment that we are solving for
a temperature map). We assume that the noise can be decomposed into a
correlated part consisting of a linear combination (described by the
matrix ${\bf F}$) of a set of baseline offsets ${\bf a}$ and a white
noise component ${\bf n}$,
\begin{equation}
{\bf n^\prime} = {\bf F}{\bf a} + {\bf n},  \label{TOD2}
\end{equation}
with covariance matrix 
\begin{equation}
\langle {\bf n^\prime} ({\bf n^\prime})^T \rangle = {\bf F} {\bf C}_a {\bf F}^T + {\bf C}_n, \qquad {\bf C}_a = \langle {\bf a}{\bf a^T} \rangle, 
\qquad {\bf C}_n = \langle {\bf n}{\bf n^T} \rangle .\label{TOD3}
\end{equation}
As shown in KKP05, if the distributions of the offsets ${\bf a}$
and white noise ${\bf n}$ are assumed to be Gaussian, the maximimum 
likelihood map and offsets are give by minimising the $\chi^2$,
\begin{equation}
\chi^2 = ({\bf y} - {\bf F}{\bf a} - {\bf P}{\bf m})^T {\bf C}_n^{-1}
({\bf y} - {\bf F}{\bf a} - {\bf P}{\bf m}) + {\bf a}^T {\bf C}_a^{-1}
{\bf a} ,\label{TOD4}
\end{equation}
giving the solutions
\beglet
\begin{eqnarray}
{\bf m} &=& ({\bf P}^T {\bf C}_n^{-1}{\bf P})^{-1} {\bf P}^T {\bf C}_n^{-1}({\bf y} - {\bf F} {\bf a}),  \label{TOD5a}  \\
{\bf a} &=& ({\bf F}^T {\bf C}_n^{-1}{\bf Z}{\bf F} + {\bf C}_a^{-1})^{-1}
{\bf F}^T{\bf C}_n^{-1} {\bf Z} {\bf y}, \label{TOD5b}
\end{eqnarray}
where
\begin{equation}
{\bf Z} = {\bf I} - {\bf P}({\bf P}^T{\bf C}_n^{-1}{\bf P})^{-1}{\bf P}^T{\bf C}_n^{-1}. \label{TOD5c} 
\end{equation}
\endlet
The Fisher matrix of the baseline offsets is given by the second 
derivative of (\ref{TOD4}) with respect to the baselines ${\bf a}$,
\begin{equation}
{\bf F}_a = ({\bf F}^T{\bf C}^{-1}_n{\bf Z}{\bf F} + {\bf C}_a^{-1}),
\label{TOD6}
\end{equation}
and we will assume that the {\it posterior} distribution of the offsets
${\bf a}$ is given by a Gaussian distribution of the form
\begin{equation}
(2 \pi \vert {\bf F}_a^{-1} \vert )^{-1/2} {\rm exp} \left ( -{1 \over 2}
{\bf a}^T {\bf F}_a {\bf a} \right ). \label{TOD7}
\end{equation}
The first term in (\ref{TOD6}) depends only on the scanning geometry
and white noise component of the detector noise. The second term
depends on the form of the detector noise spectrum at low frequencies. For
\planck -like noise properties (detector knee frequencies of $\simlt
30$mHz), the first term is dominant and represents an {\it
irreducible} contribution to the baseline offset covariance. In the
rest of this paper we will assume that the irreducible term dominates
and hence that the striping errors are independent of the exact form
of the low frequency noise spectrum. In other words, we ignore ${\bf
C}_a$ and assume white noise errors in the TOD throughout. This should
be an excellent approximation for \plancks (see Figure 5 of E04). It is
straighforward to generalise to the results of this paper include
correlations ${\bf C}_a$ in the baseline offsets. Methods for
estimating ${\bf C}_a$ from the TOD are discussed in KKP05.

The generalisation of the above analysis to polarization sensitive
detectors is straightforward. Define ${\bf y}^d$ to be the TOD for
detector $d$ with orientation  $\alpha^d$ with respect to a fixed
spherical polar coordinate system.  In terms of the Stokes parameters
$I$, $Q$ and $U$ in this fixed coordinate system, the detector response
is 
\begin{equation}
{\bf y}^d = {1 \over 2} \left ({\bf I} + {\bf Q}\cos(2\alpha^d) + 
{\bf U}\sin(2 \alpha^d) \right )  + {\bf n^\prime}, \label{TOD8}
\end{equation}
(see Figure 1 and Section 3.1 for a precise definition of the angle
$\alpha$). If the noise covariance ${\bf C}_n$ is diagonal, $C_n
\equiv \sigma_d^2$ and the pointing
matrix $P$ is consists of ones and zeros (denoting when a map pixel falls
onto an element of the TOD) the first term in equation (\ref{TOD4}) can be
written as
\begin{equation}
\tilde\chi^2 = \sum_{pd} \sum_{d i\subset p}  
{( y^d_i - (Fa)_i - \hat y^d_i)^2 \over  \sigma_d^2},
 \label{TOD9}
\end{equation}
where $p$ denotes the map pixel,  $di
\subset p$ denotes the TOD elements of detector $d$ that lie within
map pixel $p$ and $\hat y^d_i$ is the expected TOD signal for detector
$d$ from the best fitting maps, $I_p$, $Q_p$, $U_p$, derived by
minimising the $\chi^2$ of equation (\ref{TOD4}). If correlated errors
are small, the second derivative of (\ref{TOD9}) gives the inverse
covariance matrix for $I_p$, $Q_p$, $U_p$, at pixel $p$, 
\begin{eqnarray}
  {\bf M}_p = \left( \begin{array}{ccc} \sum{1 \over \sigma_d^2} &
 \sum {\cos(2 \alpha^d_i) \over \sigma_d^2} & \sum {\sin(2 \alpha^d_i)
 \over \sigma_d^2} \\ \sum {\cos(2 \alpha^d_i) \over \sigma_d^2} &
 \sum {\cos^2(2 \alpha^d_i) \over \sigma_d^2} & \sum {\cos(2
 \alpha^d_i)\sin(2 \alpha^d_i) \over \sigma_d^2} \\ \sum {\sin(2
 \alpha^d_i) \over \sigma_d^2} & \sum {\cos(2 \alpha^d_i)\sin(2
 \alpha^d_i) \over \sigma_d^2} & \sum {\sin^2(2 \alpha^d_i)
  \over \sigma_d^2} ,   \end{array} \right ),
 \label{TOD10}
\end{eqnarray}        
where the sums extend over all detector TOD elements $di \subset
p$. The condition number of the matrix $M$ is often used to determine
whether the Stokes parameters are well constrained in map pixel $p$
(see {e.g.} A06). Evidently, equation (\ref{TOD10})
applies only if the TOD noise is strictly white and in general the noise
will be correlated.  Section 4 describes a
fast Monte-Carlo method for estimating these correlated errors.

\section{Analysis of a Ring Torus}

\subsection{Ring torus geometry}

\begin{figure*}

\vskip 2.8 truein

\includegraphics{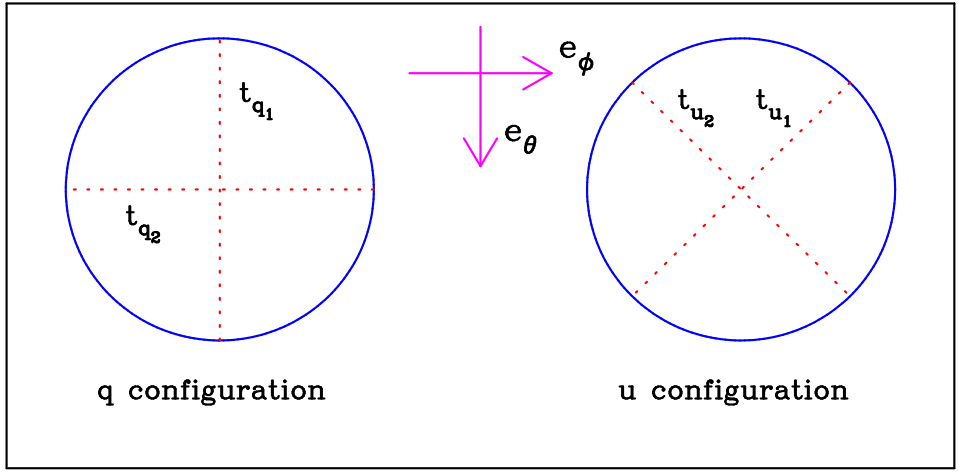}

\caption
{Polarization orientations for two pairs of bolometers in the standard
$Q$ and $U$ configurations as discussed in Section 3.1. The polarization
angles $\alpha$ for arbitrary detector orientation are defined with
respect to the unit vectors $\hat e_\phi$ and $\hat e_\theta$ defined
with respect to a reference spherical coordinate system.}

\label{figure1}

\end{figure*}

In this Section, we analyse destriping errors in the idealised case of
ring torus scanning geometry with two pairs of bolometers in the
`standard' $Q$ and $U$ configurations shown in Figure
\ref{figure1}. Each detector pair is assumed to point to the same spot
on the sky at a boresight angle of $\theta_b$ to the spin axis. The
sky is covered by a ring torus composed of a set of discrete $N_{\rm
ring} = 2 \pi/\Delta \alpha$ rings of width $\Delta \alpha$ as the spin
axis is repointed along a great circle. The detector orientations of the $Q$- 
and $U$-configurations with
respect to a fixed polarization basis on the sky can be specified by the
the angles
\begin{equation}
\alpha^q = \cos^{-1} (- \hat {\bf q}_1 \cdot \hat {\bf e}_\phi), \quad
\alpha^u = \cos^{-1} (- \hat {\bf u}_1 \cdot \hat {\bf e}_\phi) - {\pi
\over 4}. \label{M0}
\end{equation}

Since the detectors are assumed to point to the same spot on the sky,
$\alpha^q = \alpha^u$ for detectors in the standard configuration of
Figure \ref{figure1}, 
though below we will write down general results for arbitrary angles.
 The response of each of
the detectors is denoted $t^i_{q_1}$, $t^i_{q_2}$, $t^i_{u_1}$,
$t^i_{u_2}$, as shown in Figure \ref{figure1}. If the noise in each
TOD is white, with identical variance $\sigma^2_d$, unbiased $I$, $Q$,
and $U$ maps can be constructed by minimising (\ref{TOD9}):
\beglet
\begin{eqnarray} 
I_p & = & \sum_{i \subset p} {1 \over 2} (t^i_{q_1} +  t^i_{q_2} + 
 +t^i_{u_1} + t^i_{u_2}),  \label{M1a} \\
Q_p & = & {S_2^p(C_q^p - S_u^p) - S_3^p(S_q^p + C_u^p) \over
\left( S_1^pS_2^p - (S_3^p)^2 \right )},   \label{M1b} \\
U_p & = & {S_1^p(S_q^p - C_u^p) - S_3^p(C_q^p - S_u^p) \over
\left( S_1^pS_2^p - (S_3^p)^2 \right )},  \label{M1c}
\end{eqnarray}
\endlet
where
\beglet
\begin{eqnarray}
C_{\{q,u\}}^p & = & \sum_{i \subset p} (t^i_{{\{q,u\}}_1} - t^i_{{\{q,u\}}_2})
 \cos (2 \alpha^{\{q,u\}}_i),  \label{M2a} \\
S_{\{q,u\}}^p & = & \sum_{i \subset p} (t^i_{{\{q,u\}}_1} - t^i_{{\{q,u\}}_2}) 
\sin (2 \alpha^{\{q,u\}}_i),  \label{M2b} \\
S^p_1 & = & \sum_{i \subset p} (\cos^2 2 \alpha^q_i + \sin^2 2\alpha^u_i),
  \label{M2c} \\
S^p_2 & = & \sum_{i \subset p} (\sin^2 2 \alpha^q_i + \cos^2 2\alpha^u_i),  \label{M2d} \\
S^p_3 & = & \sum_{i \subset p} (\cos 2 \alpha^q_i\sin 2\alpha^q_i - \cos
 2\alpha^u_i \sin 2 \alpha^u_i).  \label{M2e}
\end{eqnarray}
\endlet

The problem that we wish to solve in this Section is as follows:
Assume that destriping is performed by fitting a single baseline
offset for each scanning ring for each detector and assume further
that the errors in the baseline offsets are uncorrelated and
charaterised by dispersions $\sigma^2_q$ and $\sigma^2_u$ for the $Q$-
and $U$-configurations respectively. We wish to calculate the biases
in the temperature and polarization power spectra associated with
these errors. This provides a good model for the effects of residual
striping errors on CMB power spectra. As we will show in the rest of this
Section, this problem can be solved analytically for a ring torus
geometry.

\begin{figure*}

\vskip 2.6 truein

\includegraphics{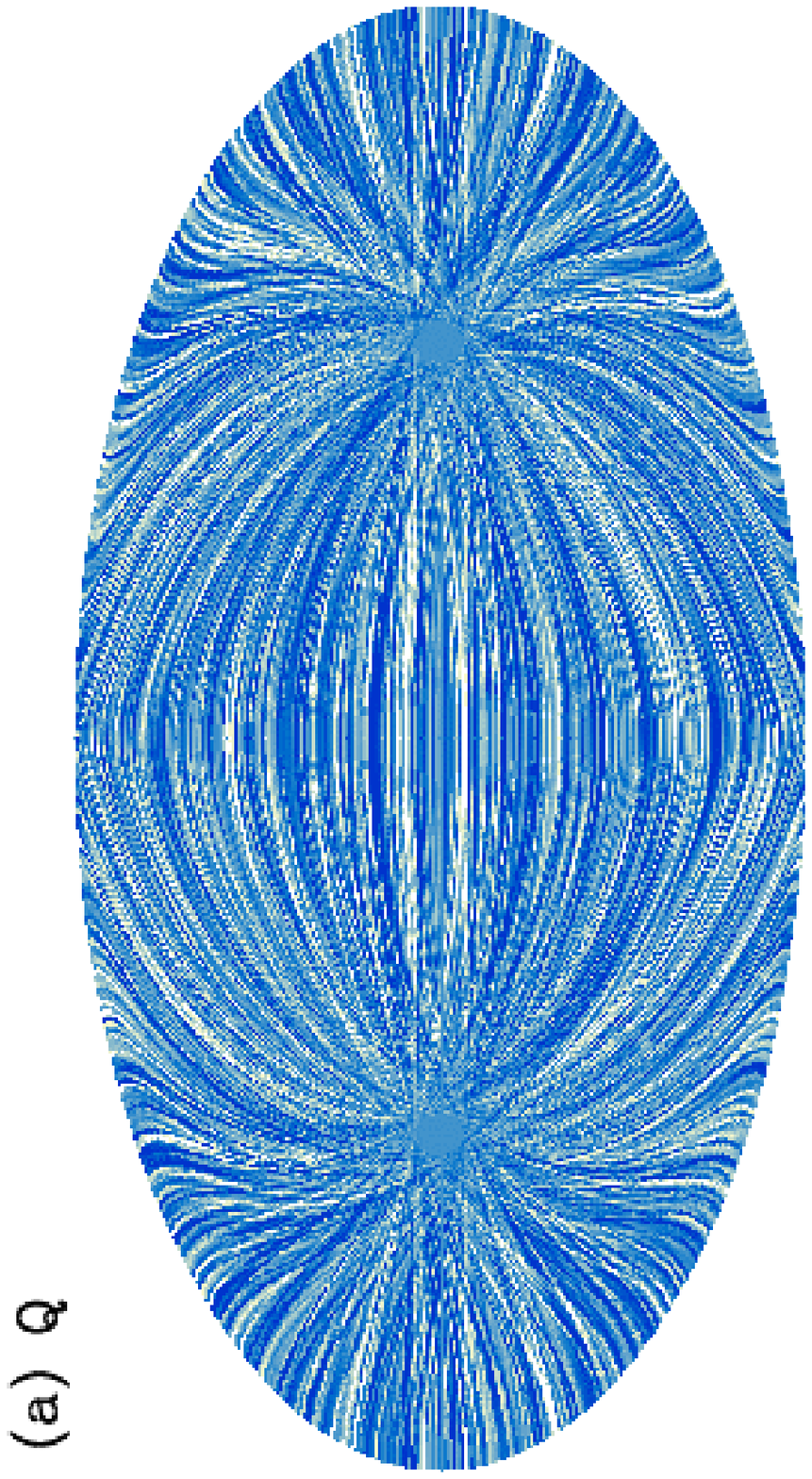}
\includegraphics{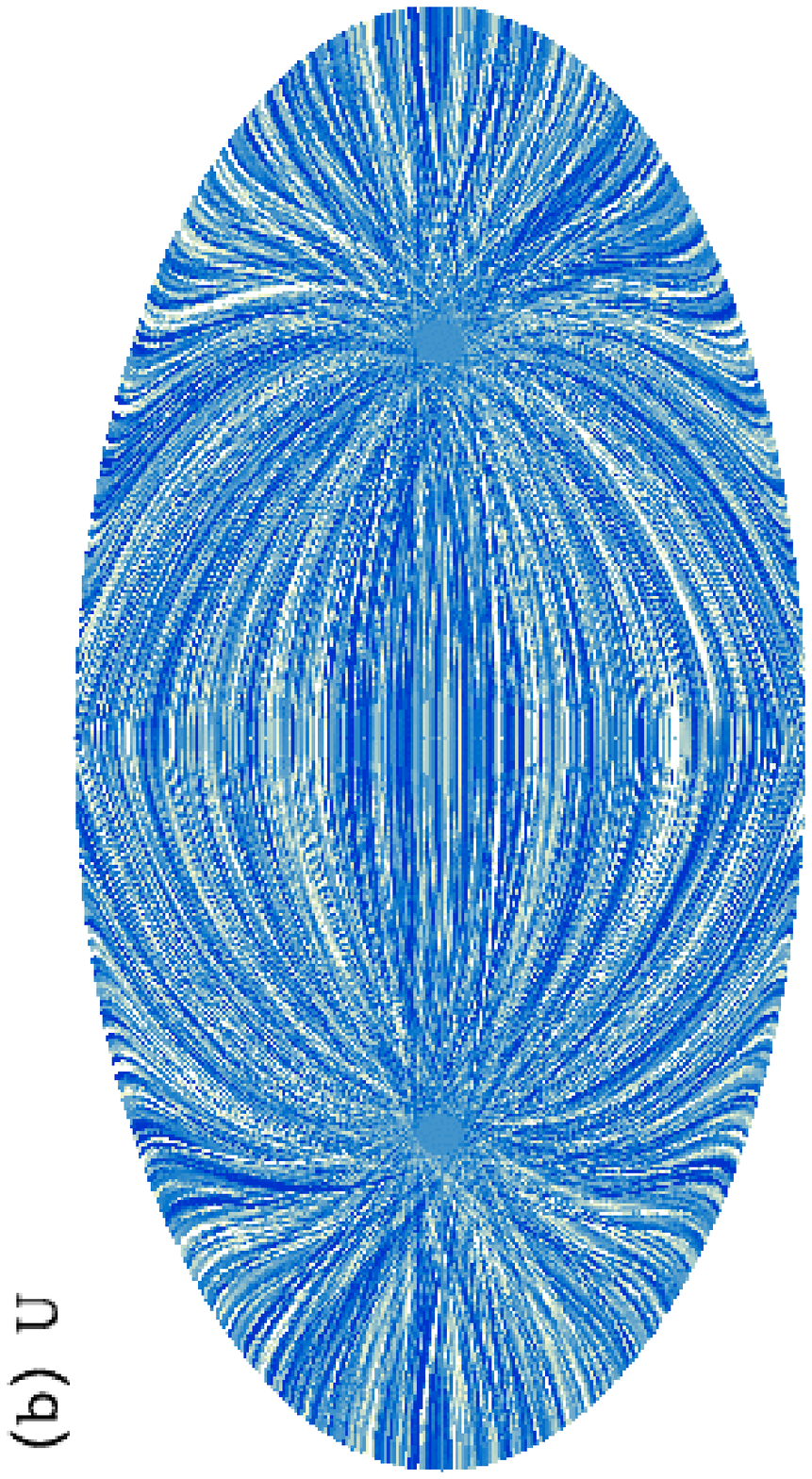}

\caption
{Q and U maps generated from random striping errors for two
polarization detector pairs in the standard configuration of Figure 1
and a ring torus scanning geometry with a boresight angle of $\theta_b
= 85^\circ$.}

\label{figure2}

\end{figure*}

 To compare with the analytic calculations, we have run a set of
$10000$ simulations constructing $I$, $Q$ and $U$ maps, using equations
(\ref{M1a}-- \ref{M1c}) for a ring torus geometry with random ring
offsets. $Q$ and $U$ maps for one such realisation are shown in Figure
\ref{figure2}. In this set of simulations, we use $N_{\rm ring}=2160$
rings of width $\Delta \alpha = 2\pi/N_{\rm ring}$ composed of $N_{\rm
ring}$ ring pixels. Maps were constructed using an igloo pixelisation
scheme (E04) with pixels of size $\Delta\theta_c =
0.25^\circ$. We will refer to these maps as `error maps'. 

The assumptions underlying these simulations are, of course, highly
idealized.  As mentioned above, with these assumptions the error power
spectra can be computed analytically. It is straightforward to
generalise these simulations to realistic \planck-like scanning
strategies and to include correlations between (an arbitrary number)
of baseline offsets. Simulations of this sort are described in Section
4. As we will show, the \plancks scanning strategy (see SPP05) is so
close to a ring torus that the model described in the next Section
provides quite a good description of the destriping errors expected
for \planck.

\subsection{Error power spectra}

The spherical harmonic transforms of the error maps can be written as
\beglet
\begin{eqnarray}
 a^T_{\ell m} &=& \sum_{ik} w_{ik}
\Omega_{ik} I_{ik}Y^*_{\ell m}( \theta_{ik}, \phi_{ik}),      \label{RT2a} \\
 a^E_{\ell m} &=& -{1 \over 2} \sum_{ik} w_{ik}
\Omega_{ik} (Q_{ik}R^{+*}_{\ell m} + i U_{ik}R^{-*}_{\ell m}),     \label{RT2b} \\
 a^B_{\ell m} &=& {i \over 2} \sum_{ik} w_{ik}
\Omega_{ik} (Q_{ik}R^{-*}_{\ell m} + i U_{ik}R^{+*}_{\ell m}),      \label{RT2c}
\end{eqnarray}
\endlet
where
\begin{equation}
R^+_{\ell m} = \;_2Y_{\ell m} +\;_{-2}Y_{\ell m}, \qquad R^-_{\ell m} = \;_2Y_{\ell m} - \;_{-2}Y_{\ell m}. \label{RT3}
\end{equation}
In these equations the index $k$ denotes the ring number, $i$ denotes the pixel
number within the ring and $\Omega_{ik}$ is the solid angle of the
ring pixel. $I_{ik}$ is the total intensity in ring $k$, and $Q_{ik}$
and $U_{ik}$ are the Stokes parameters in ring $k$ defined {\it with
respect to a fixed polarization basis}.  The weight factors $w_{ik}$
account for the averaging of the ring pixels in constructing the map
and thus are proportional to the inverse of the `hit count' distribution
for the ring torus scanning strategy.  It is useful to define the
auxiliary functions $\hat P_{\ell m}$, $\hat G^+_{\ell m}$ and $\hat
G^-_{\ell m}$: \beglet
\begin{eqnarray}
Y_{\ell m} &=& A^m_\ell P^m_\ell(\cos \theta) e^{ i m \phi} \equiv \hat P_{\ell m} e^{ i m \phi},    \label{RT4a} \\
R^{+}_{\ell m}  &=& 2 \sqrt
2 N_\ell A_\ell^m G^+_{\ell m} e^{i m \phi} \equiv \hat G^+_{\ell m} e^{ i m \phi},  \label{RT4b} \\
R^{-}_{\ell m}  &=& -2 \sqrt
2 N_\ell A_\ell^m G^-_{\ell m} e^{i m \phi} \equiv \hat G^-_{\ell m} e^{ i m \phi}, \label{RT4c}
\end{eqnarray}
\endlet
where
\beglet
\begin{eqnarray}
A_{\ell}^{m} &=& \left ( {2 \ell + 1 \over 4 \pi} {( \ell - m)! \over
(\ell + m)!}  \right )^{1/2},     \label{RT5a} \\
N_{\ell} &=& \left ( {2 (\ell - 2)! \over (\ell + 2) !}  \right )^{1/2}, \label{RT5b}
\end{eqnarray}
\endlet
and the functions $G^+_{\ell m}$ and $G^{-}_{\ell m}$ are
\beglet
\begin{eqnarray}
G^+_{\ell m} & = &
 - \left ( {\ell - m^2 \over \sin^2 \theta} + { 1 \over 2}
\ell(\ell - 1) \right ) P_\ell^m(\cos \theta) + (\ell +m) {\cos \theta
\over \sin^2 \theta} P_{\ell -1}^m (\cos \theta),  \label{RT6a} \\
G^-_{\ell m}  & = &   {m \over \sin^2 \theta}
\left \{  (\ell - 1) \cos \theta  P_\ell^m(\cos \theta) - (\ell +m) 
 P_{\ell -1}^m (\cos \theta) \right \}  .\label{RT6b}
\end{eqnarray}
\endlet 
(See, for example, Kamionkowski, Kosowsky and Stebbins, 1997,
and references therein.) Note that the functions $\hat P$, $\hat G^+$
and $\hat G^-$ satisfy
\begin{equation}
\hat P_{\ell - m} = (-1)^m \hat P_{\ell m}, \quad \hat G^+_{\ell - m} = (-1)^m
\hat G^+_{\ell m}, \quad \hat G^-_{\ell -m } = -(-1)^m \hat G^-_{\ell m}.
\end{equation}
To compute the power spectra of the error maps, reorient each ring to
a new coordinate system $(\theta^\prime, \phi^\prime)$ in which the
spin axis is aligned with the new $z^\prime$ axis. The weight function
$w_{ik}$ is normalised so that
\begin{equation}
\sum_{ik} w_{ik} \Omega_{ik} = 4 \pi \sin \theta_b,  \label{RT7}
\end{equation}
hence in the new coordinate system the weight factor for each ring is given by
\begin{equation}
 w_{ik} = {1 \over 2} \sin \theta_b \vert \sin \phi^\prime_{ik} \vert.  \label{RT8}
\end{equation}
Define the integral $I(m)$ to be
\begin{equation}
I(m) =  \int_0^{2 \pi} \vert \sin \phi^\prime \vert 
{\rm e}^{-i m \phi^\prime} d\phi^\prime = {2 \over (1 - m^2)} ( 1 + (-1)^m),  \label{RT9}
\end{equation}
then the contribution of a single ring $k$ to the $T$, $E$ and $B$
mode harmonic coefficients is
\beglet
\begin{eqnarray}
 a^{\prime Tk}_{\ell m}  &=& {1 \over 2} \sin \theta_b I_k 
  \Delta \alpha I(m) \hat P_{\ell}^{m}(\cos \theta_b) ,    \label{RT10a} \\
 a^{\prime Ek}_{\ell m}  &=& -{1 \over 4} \sin \theta_b \Delta \alpha I(m) 
\left [ Q_k \hat G^{+}_{\ell m}(\cos \theta_b) + i U_k \hat G^{-}_{\ell m} (\cos \theta_b) \right ],    \label{RT10b} \\
 a^{\prime Bk}_{\ell m}  &=& \;\;\;{i \over 4} \sin \theta_b \Delta \alpha I(m) 
\left [ Q_k \hat G^{-}_{\ell m}(\cos \theta_b) + i U_k \hat G^{+}_{\ell m} (\cos \theta_b) \right ],   \label{RT10c}
\end{eqnarray}
\endlet
where 
\beglet
\begin{eqnarray}
I_k &=& {1 \over 2} (t^k_{q_1} + t^k_{q_2} + t^k_{u_1} + t^k_{u_2}),
 \label{D1a}  \\
Q_k &=& t^k_{q_1} - t^k_{q_2}, \label{D1b} \\
U_k &=& t^k_{u_1} - t^k_{u_2}, \label{D1c} 
\end{eqnarray}
\endlet
and the quantities $t^k_{q_1}$ {\it
etc} are the random numbers assigned to each detector ring in the
standard configuration shown in Figure 1. The transformation to the
new coordinate system thus leads to a simple relationship between the 
$E$ and $B$ mode harmonic coefficients and the quantities recorded 
by each detector.
Evidently if the baseline offsets are uncorrelated,
\beglet
\begin{eqnarray}
 \langle I^2 _k \rangle &=& {1 \over 4} \left ( \langle t^2_{q_1} \rangle
+ \langle t^2_{q_2} \rangle +  \langle t^2_{u_1} \rangle + \langle t^2_{u_2}  \rangle
\right ),    \label{RT11a} \\
 \langle Q^2 _k \rangle &=& \langle t^2_{q_1} \rangle
+ \langle t^2_{q_2} \rangle,    \label{RT11b} \\
 \langle U^2 _k \rangle &=& \langle t^2_{u_1} \rangle
+ \langle t^2_{u_2} \rangle,     \label{RT11c} \\
 \langle I_kQ_k \rangle &=& {1 \over 2} (\langle t^2_{q_1} \rangle
- \langle t^2_{q_2} \rangle),     \label{RT11d} \\
 \langle I_kU_k \rangle &=&  {1 \over 2} (\langle t^2_{u_1} \rangle
- \langle t^2_{u_2} \rangle).   \label{RT11e}
\end{eqnarray}
\endlet In the simulations described in the previous Section we
 assumed $\langle t^2_{q_1} \rangle = \langle t^2_{q_2} \rangle =
 \sigma^2_q$, $\langle t^2_{u_1} \rangle = \langle t^2_{u_2} \rangle =
 \sigma^2_u$, with $\sigma_q = \sigma_u = 13.6 m{\rm K}$ (though the
 actual value is unimportant for this discussion).  The spherical and
 tensorial harmonics can be rotated  to an arbitrary spherical
 polar coordinate system using the Wigner matrices $D^\ell_{m^\prime
 m} (\alpha, \beta, \gamma)$, where $\alpha$, $\beta$ and $\gamma$ are
 the Euler angles specifying the rotation (see {\it e.g.} Brink and
 Satchler 1993; Varshalovich, Moskalev and Khersonskii 1988). Thus,
each of the ring harmonic coefficients in  equations (\ref{RT10a})--(\ref{RT10c}) can be rotated back to the original coordinate system and summed to
compute the harmonic coefficients of equations  (\ref{RT2a})--(\ref{RT2c}):
\begin{equation}
a_{\ell m} = \sum _k \sum_{m^\prime} a^{\prime k}_{\ell m^\prime}
D^\ell_{ m^\prime m} ( \alpha_k, \beta_k, \gamma_k) = \sum _k
\sum_{m^\prime}a^{\prime k}_{\ell m^\prime} {\rm e}^{-im^\prime
\alpha_k} d^\ell_{m^\prime m}(\beta_k) {\rm e}^{-im \gamma_k},
\label{RT12}
\end{equation}
where the $d^\ell_{m^\prime m}$ are the reduced rotation matrices. Thus to compute the error power spectra, we set $\alpha_k=0$, $\beta_k = \pi/2$ and evaluate the integrals over $\gamma_k$. This gives
\beglet
\begin{eqnarray}
\tilde {C}^T_{\ell} &=& {\pi \over 2(2 \ell + 1)} {(\sigma_q^2 + \sigma^2_u) \over 2}  \Delta \alpha
\sin^2 \theta_b \sum_{m} I^2(m) \left ( \hat P_\ell^m ( \cos \theta_b)
\right)^2,  \label{RT12a} \\
\tilde {C}^E_{\ell} &=& { \pi \over 4(2 \ell + 1)}    \Delta \alpha
\sin^2 \theta_b \sum_{m} I^2(m)  \left [ 
\sigma^2_q (\hat G^+_{\ell m} ( \cos \theta_b))^2 + 
\sigma^2_u (\hat G^-_{\ell m} ( \cos \theta_b))^2 
\right]^2,  \label{RT12b} \\
\tilde {C}^B_{\ell} &=& { \pi \over 4 (2 \ell + 1)}    \Delta \alpha
\sin^2 \theta_b \sum_{m} I^2(m)  \left [ 
\sigma^2_q (\hat G^-_{\ell m} ( \cos \theta_b))^2 + 
\sigma^2_u (\hat G^+_{\ell m} ( \cos \theta_b))^2 
\right]^2,  \label{RT12c} \\
\tilde {C}^X_{\ell} &=& -{\pi \over  4 (2 \ell + 1)}    \Delta \alpha
\sin^2 \theta_b {(\langle t^2_{q_1} \rangle - \langle t^2_{q_2} \rangle) 
\over 2}
\sum_{m} I^2(m) \hat P_\ell^m(\rm \cos \theta_b)
\hat G^+_{\ell m} ( \cos \theta_b). \label{RT12d}
\end{eqnarray}
\endlet
Equation (\ref{RT12a}) is identical to  equation (19) of E05. Equations
(\ref{RT12b}--\ref{RT12d}) generalise the results of E05 to polarization.

\begin{figure*}

\vskip 5.3 truein

\includegraphics{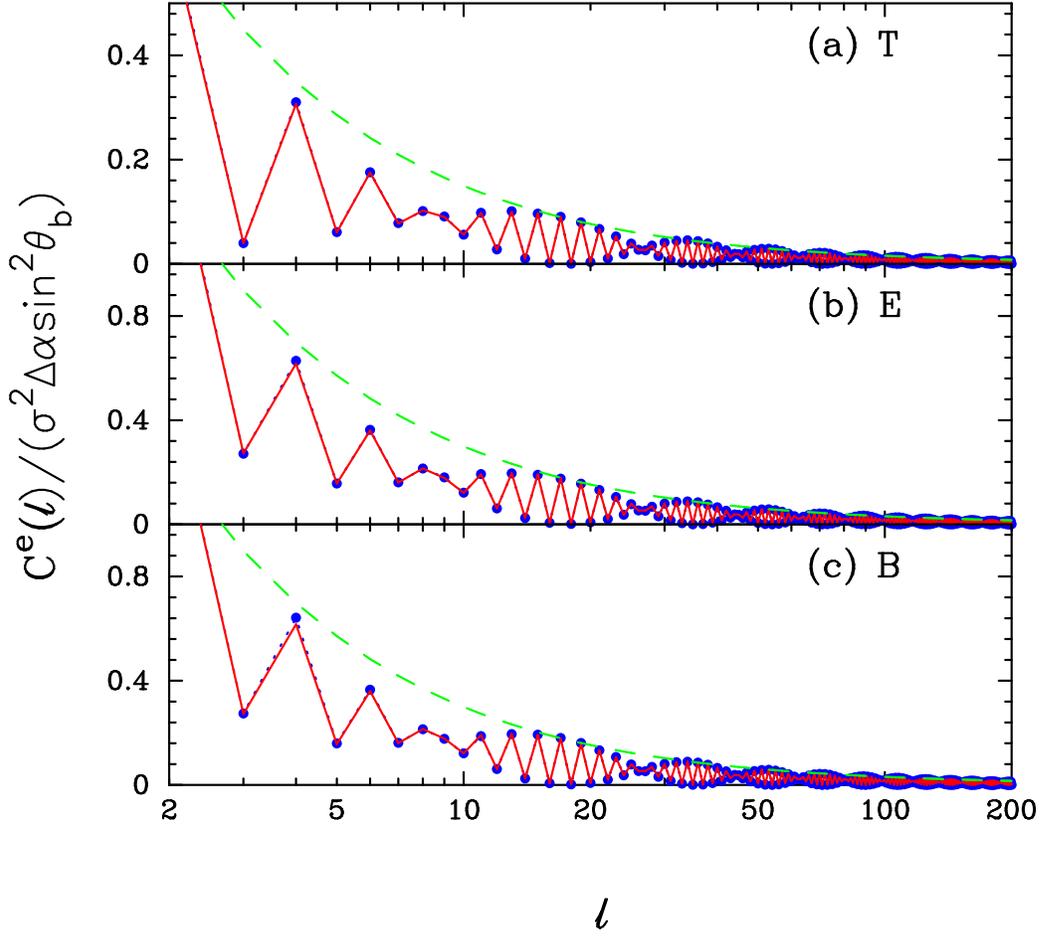}

\caption
{Destriping power spectrum errors for the ring torus geometry analysed
in the text. A boresight angle of $85^\circ$ has been assumed. The
upper panel (a) shows the results for the temperature power spectrum ,
the central panel (b) shows results for the polarization E-mode power
spectrum and the lower panel (c) shows results for the B-mode.  In each
case, the (blue) points show the results from $10^4$ simulations with
parameters as discussed in the text. The (red) solid lines show the
analytic predictions of equations (\ref{RT12a})- (\ref{RT12c}). The
(green) dashed lines show the high $\ell$ envelope $\pi/(2 \ell +1)$
for $T$ and $2 \pi/(2\ell + 1)$ for $E$ and $B$.}

\label{figure3}

\end{figure*}

Some features of these results (useful for understanding real
experiments) are worth pointing out. If $\sigma^2_q = \sigma^2_u$, the
error power spectra of the $E$ and $B$ modes are
identical. Furthermore, the dominant terms in both power spectra are
the $\hat G^+_{\ell m}$ terms which are similar to the $\hat P_\ell^m$
terms at high multipoles. Thus, we expect the error power spectra in
the $E$ and $B$ modes to have about twice the amplitude of the
temperature error power spectrum. The ring variances $\sigma^2_q$ and
$\sigma^2_u$ are determined largely by the white noise in the TOD and
the scanning geometry (equation \ref{TOD6}), and so differences in the
detector noise levels will directly effect the destriping error
contributions to the power spectra. Finally, if $\langle t^2_{q_1}
\rangle = \langle t^2_{q_2} \rangle$, equation (\ref{RT12d}) show that
destriping errors on the TE (denoted X in this paper) power spectrum
will be identically zero, even if the destriping errors are large.

Figure \ref{figure3} compares equations (\ref{RT12a})--(\ref{RT12c})
with the average power spectra determined from $10^4$ simulations as
described in Section 3.1 (see Figure \ref{figure1}). (We do not plot
$\tilde C^X_\ell$ since the simulations assume identical ring
variances for each detector and so the error power spectrum averages
to zero. The dashed lines in Figure \ref{figure3} show the envelopes
$\tilde C^e_\ell \propto 1/(2 \ell + 1)$, valid in the limit $\ell \gg
1$ for a boresight angle $\theta_b = \pi/2$ (E04). Evidently, the
simulations and theory are in perfect agreement. The `sawtooth'
pattern, with a $1/\ell$, envelope is a characteristic feature of map
making in a total power experiment with a near ring torus scanning
geometry.  This pattern can be seen in, for example, the analysis of \plancks
simulations described by A06. It is worth emphasing the point made in
the introduction that destriping errors for a \planck-like experiment
cannot be removed by applying an `optimal' least-squares map making
code, since it is not possible to remove the `irreducible' striping
errors fixed by the detector white noise and scanning. Furthermore, if
the detector $1/f$ knee frequencies are comparable to,  or less than, the
spin frequency, simple destriping with a {\it small} number of baselines
coefficients will be sufficient to get close to the irreducible
striping errors (see {\it e.g.} E05, KKP05, A06). The analysis
presented in this Section shows that for a \planck-like experiment,
the effects of residual striping errors in the $T$, $E$ and $B$ power
spectra will be `pinned' to the detector white noise level (since this
largely fixes the variance of the baseline offsets) and will always dominate
over the white noise level at low multipoles.

\subsection{Pixel noise covariances}

Optimal power spectrum estimators based on maps require the
pixel-pixel noise covariance matrices. In the hybrid power spectrum
approach discussed in E04 and E06, PCL estimators are applied at high
multipoles and quadratic maximum likelihood (QML) estimators ({\it
e.g.} Tegmark 1997c; Tegmark and de Oliveira-Costa 2001) are applied
on degraded resolution maps to determine the power spectra at low
multipoles. In the examples discussed in E04 and E06, noise was either
ignored for the QML estimates, or assumed to be uncorrelated.
The results of the previous Section show that in a realistic
\planck-like experiment, destriping errors will dominate over white
noise at low multipoles and hence must be quantified to obtain
accurate estimates of the power spectra and their statistical
distributions. The same is true for any map-based
power spectrum estimator, including map-based Gibbs sampling methods
(see {\it e.g.} Larson \etals 2006).

For a \planck-size map, containing typically $\sim 2 \times 10^7$
pixels it is not feasible either to compute or store a full
pixel-pixel covariance matrix. However, since destriping errors are
important only at low multipoles, all that is required for power
spectrum estimation is an accurate model for the pixel-pixel covariance
matrices in low resolution maps.  In fact, for the ring torus
geometery described in the previous Section, it is possible to
calculate these covariance matrices analytically.

The power spectra computed in the previous Section are, of course,
invariant with respect to the orientation of the spherical coordinate
system, whereas the Stokes parameters $Q$ and $U$ and their associated
pixel covariance matrices are not. In the ring torus scanning
geometry, the spin axis defines a fixed plane in the sky (the ecliptic
plane for a \planck-type experiment). A common choise is to pick the
$z$-axis of the spherical coordinate system to be perpendicular to
this plane. We therefore rotate the maps from the coordinate system
shown in Figure \ref{figure2} to a new coordinate systems where the
empty holes in the maps lie at the `ecliptic' poles.

\begin{figure*}

\vskip 8.0 truein

\includegraphics{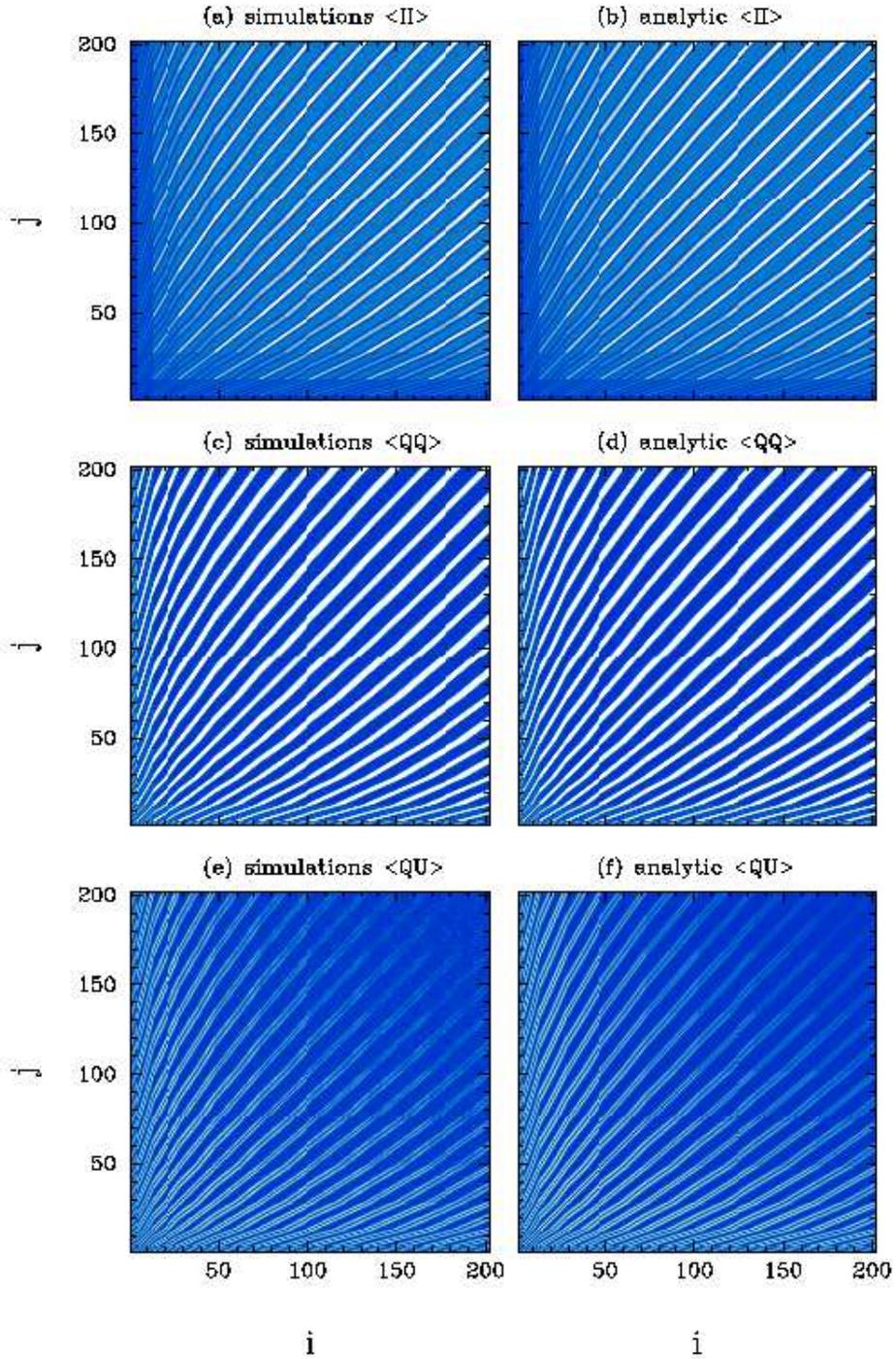}

\caption
{Pixel-pixel covariance matrices for reduced resolution ($\Delta
\theta_c = 10^\circ$ pixels) maps for a ring torus geometry with a
boresight angle $\theta_b = 85^\circ$. Since the covariance matrices
are four-fold symmetric, we show just the lower quadrant of $202\times
202$ elements. The figures to the left show the results of $10^4$
simulations (the same set of simulations used to construct the power
spectra shown in Figure \ref{figure3}) for the $II$ covariance (panel
(a)), $QQ$ covariance (panel (b)) and $QU$ cross-covariance (panel
(c)).  (The $UU$ covariance looks identical to the $QQ$ covariance.)
The figures to the right show the analytic expressions of equations
(\ref{PN2a})-- (\ref{PN2c}).}

\label{figure4}

\end{figure*}

We assume that the low resolution maps are constructed by first
convolving the high resolution maps with a Gaussian of width
$\theta_s$ and repixelising to lower resolution.
This is equivalent to multiplying the rotated harmonic
coefficients (\ref{RT2a})-- (\ref{RT2c}) by
\begin{equation}
f_\ell = {\rm exp} \left ( - {1 \over 2} \ell(\ell+1)\theta_s^2 \right), \label{Beam1}
\end{equation}
and synthesising new maps. Define the auxilliary functions
\beglet
\begin{eqnarray}
\hat \Phi_{\ell m} &=& \sum_{m^\prime} \hat P_{\ell m^\prime} (\cos \theta_b) 
d^\ell_{m^\prime m} (\pi/2) I(m^\prime),  \label{PN1a} \\
\hat \Gamma^+_{\ell m} &=& \sum_{m^\prime} \hat G^+_{\ell m^\prime} (\cos \theta_b) 
d^\ell_{m^\prime m} (\pi/2) I(m^\prime),  \label{PN1b} \\
\hat \Gamma^-_{\ell m} &=& \sum_{m^\prime} \hat G^-_{\ell m^\prime} (\cos \theta_b) 
d^\ell_{m^\prime m} (\pi/2) I(m^\prime).  \label{PN1c}
\end{eqnarray}
\endlet
After some algebra, we can compute the pixel-pixel covariance matrices of the 
synthesised maps:
\beglet
\begin{eqnarray}
 \langle I_i I_j \rangle &=&  {\pi \over 2} \sigma^2 \sin^2 \theta_b \Delta \alpha 
\sum_{\ell_1 \ell_2} \sum_{m}
  \hat P_{\ell_1 m}(\cos \theta_i)
\hat P_{\ell_2  m} (\cos \theta_j)
 \hat \Phi_{\ell_1 m} \hat \Phi_{\ell_2 -m} f_{\ell_1} f_{\ell_2} \cos(m(\phi_i - \phi_j)),   \label{PN2a} \\
 \langle Q_i Q_j \rangle &=&  {\pi \over 16} \sigma^2 \sin^2 \theta_b \Delta \alpha 
\sum_{\ell_1 \ell_2} \sum_{m} \big [
 \left (\hat  \Gamma^+_{\ell_1 m} \hat \Gamma^+_{\ell_2 -m}
- \hat \Gamma^-_{\ell_1 m} \hat \Gamma^-_{\ell_2 -m} \right)
\left (\hat G^+_{\ell_1 m}(i) \hat G^+_{\ell_2 -m}(j) - \hat G^-_{\ell_1 m}(i)
 \hat  G^-_{\ell_2 -m}(j)  \right) +  \\ \nonumber
& & \qquad \qquad \left (\hat  \Gamma^-_{\ell_1 m} \hat \Gamma^+_{\ell_2 -m}
- \hat \Gamma^+_{\ell_1 m} \hat \Gamma^-_{\ell_2 -m} \right )
\left (\hat G^-_{\ell_1 m}(i) \hat G^+_{\ell_2 -m}(j) - \hat G^+_{\ell_1 m}(i) \hat 
G^-_{\ell_2 -m}(j)  \right )
\big   ] f_{\ell_1} f_{\ell_2} \cos(m(\phi_i - \phi_j)),   \label{PN2b} \\
 \langle Q_i U_j \rangle &=&  {\pi \over 16} \sigma^2 \sin^2 \theta_b \Delta \alpha 
\sum_{\ell_1 \ell_2} \sum_{m} \big [
 - \left (\hat  \Gamma^+_{\ell_1 m} \hat \Gamma^+_{\ell_2 -m}
- \hat \Gamma^-_{\ell_1 m} \hat \Gamma^-_{\ell_2 -m} \right)
\left (\hat G^+_{\ell_1 m}(i) \hat G^-_{\ell_2 -m}(j) - \hat G^-_{\ell_1 m}(i)
 \hat  G^+_{\ell_2 -m}(j)  \right) +  \\ \nonumber
& & \qquad \qquad \left (\hat  \Gamma^+_{\ell_1 m} \hat \Gamma^-_{\ell_2 -m}
- \hat \Gamma^-_{\ell_1 m} \hat \Gamma^+_{\ell_2 -m} \right )
\left (\hat G^+_{\ell_1 m}(i) \hat G^+_{\ell_2 -m}(j) - \hat G^-_{\ell_1 m}(i) \hat 
G^-_{\ell_2 -m}(j)  \right )
\big   ] f_{\ell_1} f_{\ell_2} \sin(m(\phi_i - \phi_j)),   \label{PN2c}
\end{eqnarray}
\endlet 
where $\theta_i$, $\phi_i$ are the angular coordinates of
pixel $i$. For simplicity, we have assumed that $\sigma^2_q = \sigma^2_u = \sigma^2$, thus $\langle U_iU_j\rangle = \langle Q_iQ_j \rangle$.

Figure \ref{figure4} compares the analytic expressions of equations
(\ref{PN2a}) -(\ref{PN2c}) with the results of numerical
simulations. The simulation results are based on the same set of
simulations used to generate Figure \ref{figure3}, but degraded to a
pixel size of $\Delta\theta_c =10^\circ$ after smoothing with
a Gaussian of width $\theta_s = 8.5^\circ$. At this resolution a full
map consists of only $404$ pixels and because of the four-fold
symmetry of the covariance matrices, we plot only the lower quadrant
(pixel indices are ordered as $i = j^\prime +
N_\phi^\prime(i^\prime)(i^\prime-1)$ where the index $i^\prime$ labels
the polar angle $\theta$, with $i^\prime=1$ corresponding to the
ecliptic pole, $j^\prime$ labels the azimuthal angle $\phi^\prime$ and
$N_\phi(i^\prime)$ denotes the number of azimuthal angle bins at polar
angle $i^\prime$). The analytic results are in perfect agreement with
the numerical simulations. The general behaviour of these pixel-pixel
covariances is straightforward to understand. The pixel noise is
highly correlated along the scanning rings, almost great circles at
this resolution, modulated by the hit count distribution. The $\langle
QU \rangle$ cross-covariances are considerably smaller than the
$\langle QQ \rangle$ and $\langle UU \rangle$ covariances and could
almost certainly be ignored for most  purposes.

\section{More Realistic Simulations}

The model described in the previous Section is useful for gaining
physical insight into the effects of destriping errors in CMB
experiments. However, the focal plane and scanning geometry are highly
idealised and the baseline offsets are assumed to be strictly uncorrelated.
The  model offers, at best, a first approximation
to the effects of destriping errors in a realistic \planck-type experiment. In
this Section, we discuss a simple, fast,  and practical Monte-Carlo method 
for estimation pixel noise covariances for a more realistic experimental
configuration.

The focal plane geometry is modelled by the four \plancks $143$ GHz
polarization sensitive bolometer pairs (see {\it e.g.}  Figure 3 of
Delabrouille and Kaplan 2002) . This consists of two pairs of
bolometers in the $Q$-configuration of Figure \ref{figure1} and two
pairs in the $U$-configuration aligned along the scan direction. We assume
a scanning pattern in which the spin axis scans along the
ecliptic plane but with a slow precession of $5^\circ \sin(2
\phi)$. The focal plane and scanning geometry therefore violate
(mildly) the assumptions of the ring torus model described in the
previous Section.

\begin{figure*}
\vskip 3.8 truein

\includegraphics{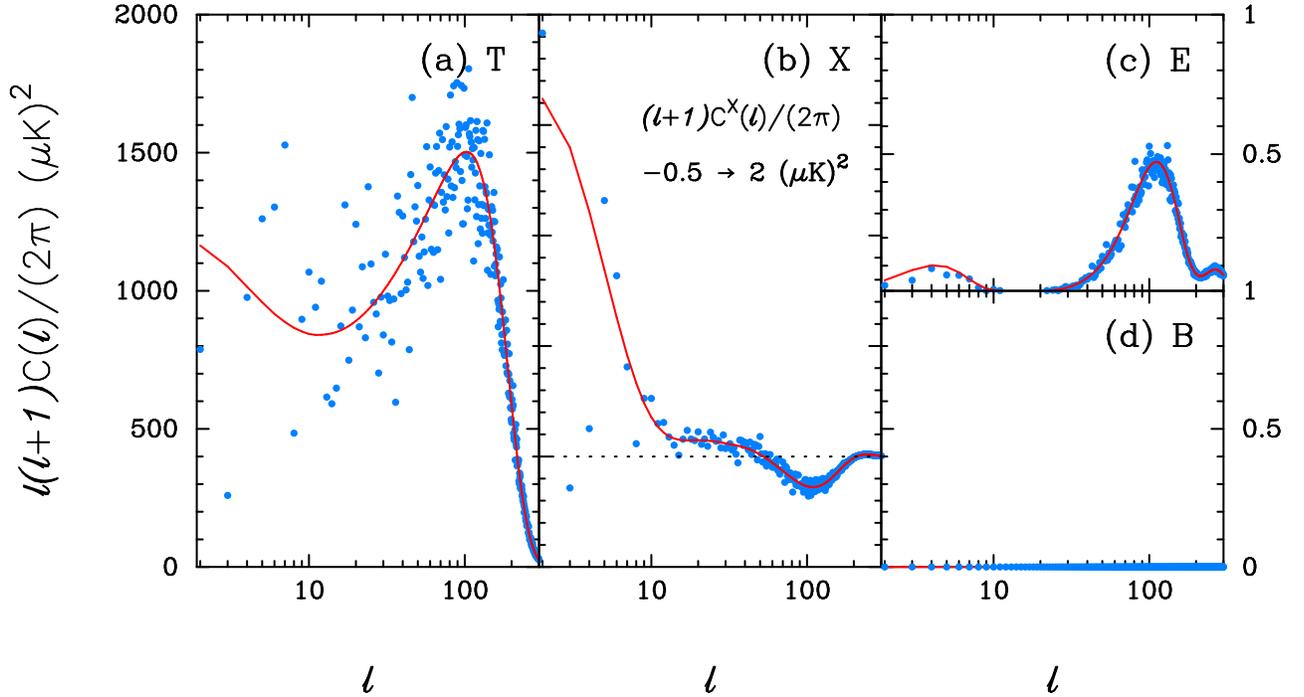}

\caption
{The (blue) points show $T$, $X$, $E$ and $B$ power spectra of a
single Gaussian realisation of a $\Lambda$CDM model with WMAP3 parameters (see
text) at a resolution of $\theta_{\rm FWHM} = 1^\circ$. The tensor 
amplitude in this model is set to zero. The (red) lines show the
expectation values of the power spectra for this $\Lambda$CDM model. 
The input $I$, $Q$ and $U$ maps for this realisation are shown in the left hand panels
of Figure \ref{figure7}.}

\label{figure5}

\end{figure*}

The input CMB maps are Gaussian realisations of a spatially flat
$\Lambda$-dominated CDM universe with the WMAP3 parameters
($\Omega_bh^2=0.0223$, $\Omega_mh^2 = 0.127$, $h=0.73$, $n_s = 0.95$,
$\tau=0.090$, $\sigma_8=0.74$, in the notation of Spergel \etals
2006). For these simulations the $B$-mode anistropy is set to zero,
{\it i.e.} the tensor mode amplitude is assumed to be negligible and
the effects of gravitational lensing (see {\it e.g.} Lewis and
Challinor, 2006, and references therein) are ignored. A Gaussian
smoothing with $\theta_s=0.425^\circ$ was applied and maps were
generated with a pixel size of $\Delta \theta_c = 0.5^\circ$. The
input power spectra for one such (noise free) realisation are shown in
Figure \ref{figure5} and the $I$, $Q$ and $U$ maps for this
realisation are shown in the left hand panels of Figure \ref{figure7}.
TOD were generated for each of the eight bolometers in $1080$ scanning
rings consisting of $1080$ ring pixels. The scanning rings partially
overlap with a width of $30^\prime$ and a separation of
$20^\prime$. Only Gaussian white noise was included in the TOD (with an
{\it rms} of $54\mu$K per ring pixel) so the simulations are designed
to compute the irreducible white noise contribution to the destriping
errors discussed in Section 2. Destriping was performed by determining
one baseline coefficient for each detector scanning ring. This is done
by minimising for each detector
\begin{equation}
\tilde\chi^2 = \sum_p \sum_{ijkk^\prime\subset p} {1 \over 2
\sigma^2 n_p} ( y^{ik} - a_k - \hat{q_i} - y^{jk^\prime} +
a_{k^\prime} + \hat{q_j})^2 + \lambda \left ( \sum_k a_k \right)^2
\label{Sim1}
\end{equation}
where $n_p$ is the hit count in map pixel $p$ and $k$ denotes the
scanning ring. The second term enforces the condition  $\sum a_k =
0$. Provided $\lambda$ is chosen to be large enough, the solutions
for the $a_k$ (and their errors) are independent of $\lambda$. The
correction terms $\hat q_i$ and $\hat q_j$ in (\ref{Sim1}) correct for
the small {\it orientation dependent} polarization contribution to the
TOD (equation \ref{TOD8}) determined from the $Q$ and $U$ maps
constructed from all of the detectors.  (Maps from the destriped noisy
TODs were constructed by a straightforward generalisation of equations
(\ref{M1a})-- (\ref{M1c}) to eight detectors.) The solution for the
offsets $a_k$ is therefore iterative: offsets are determined by first
minimising (\ref{Sim1}) with $\hat{q_i}$ set to zero, constructing
$I$, $Q$ and $U$ maps using these offsets and then recomputing offsets
by using the $Q$ and $U$ maps to compute the polarization corrections
$\hat q_i$. The solutions converge very rapidly to a stable solution.
The first term in the $\chi^2$ of equation (\ref{Sim1}) is an
approximation to 
\begin{equation}
\chi^2 = ({\bf y} - {\bf F}{\bf a})^T{\bf Z}^T {\bf C}_n^{-1} {\bf Z}
({\bf y} - {\bf F}{\bf a}), \label{Sim2}
\end{equation}
derived by substituting the solution (\ref{TOD5a}) back into equation (\ref{TOD4}).

\begin{figure*}

\vskip 3.8 truein

\includegraphics{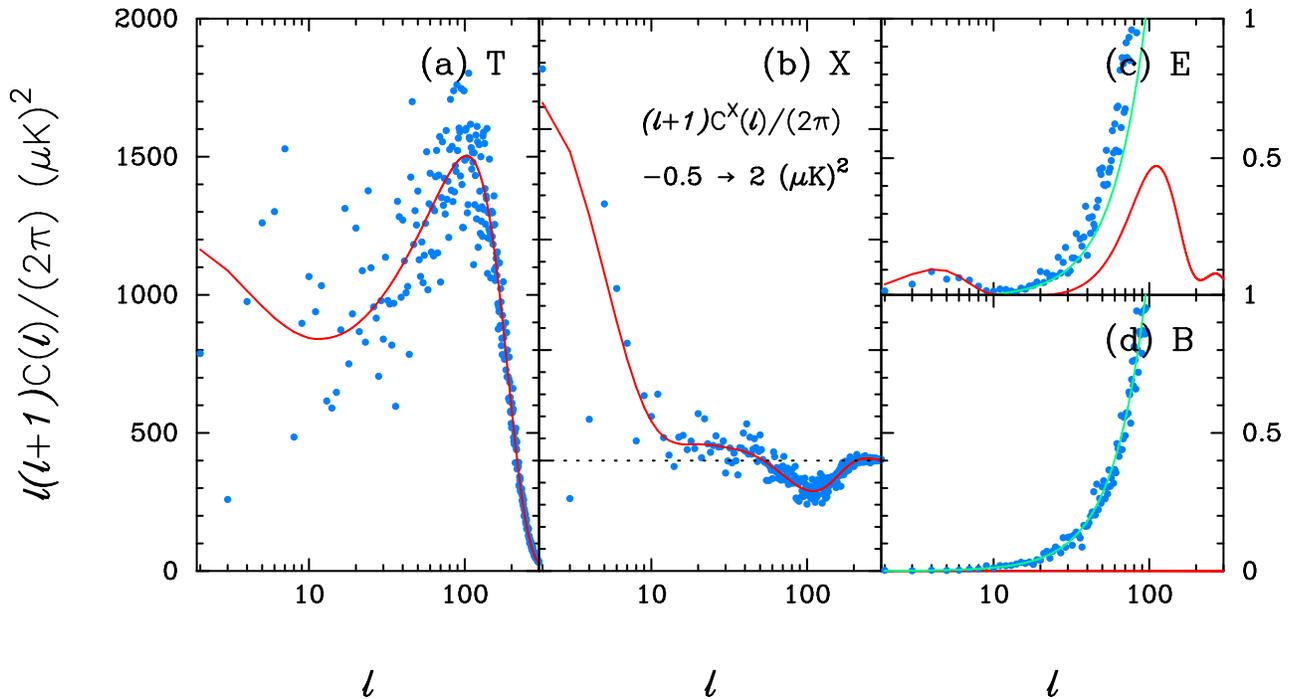}

\caption
{The (blue) points show CMB power spectrum of a noisy
 realisation of the $\Lambda$CDM model shown in Figure 5, but 
after polarization destriping as described in the text. The noise
parameters, focal plane geometry and scanning strategy have been
chosen to match the \plancks $143$ GHz polarization sensitive bolometers,
but with the noise renormalised to compensate for the degraded resolution
of the simulations. The (red) lines show the
expectation values of the power spectra for this $\Lambda$CDM model
as plotted in Figure 5. The (green) lines in panels (c) and (d) show
the white noise errors on the polarisation power spectra. }

\label{figure6}

\end{figure*}

The white noise level applied to the TOD was chosen so that the final
maps have a white noise amplitude in each $30^\prime \times 30^\prime$
map pixel about equal to that expected for the $8$ polarized \plancks
$143$GHz detectors. The simulations should thus give a good indication
of what we might expect from \plancks at low multipoles.
The resolution of the simulation (set by $1080$ scanning rings,
compared to the $8640$ scanning rings for \planck) has been chosen so
that it is possible to generate large numbers of 
simulations of the TODs on a workstation rather than a supercomputer. It is
straightforward to generalise the simulations described in this
Section to full \plancks resolution and to relax the assumptions of
white noise in the TOD and one baseline per detector ring.

\begin{figure*}

\vskip 7.1 truein

\includegraphics{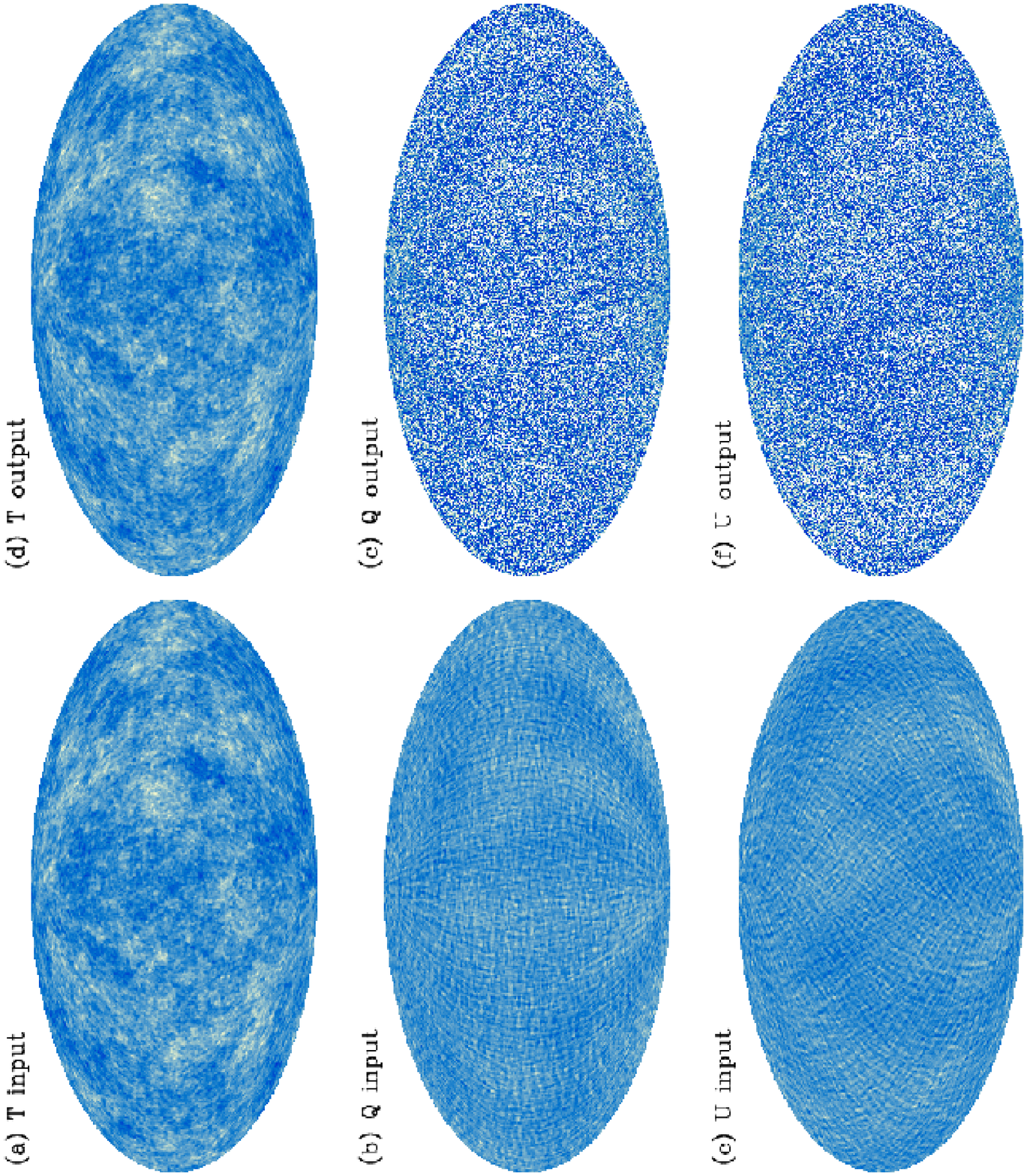}

\caption
{The pictures to the left (a)-(c) show the noise free input $I$, $Q$
and $U$ maps used to generate the power spectra plotted in Figure
\ref{figure5}.  The pictures to the right (d)-(f) show the output maps
with realistic instrument noise and polarization destriping
corresponding to the power spectra plotted in Figure
\ref{figure6}. Note that the $Q$ and $U$ maps are dominated by white
noise, though some signal from low multipoles can be discerned.}

\label{figure7}

\end{figure*}

\begin{figure*}

\vskip 7.1 truein

\includegraphics{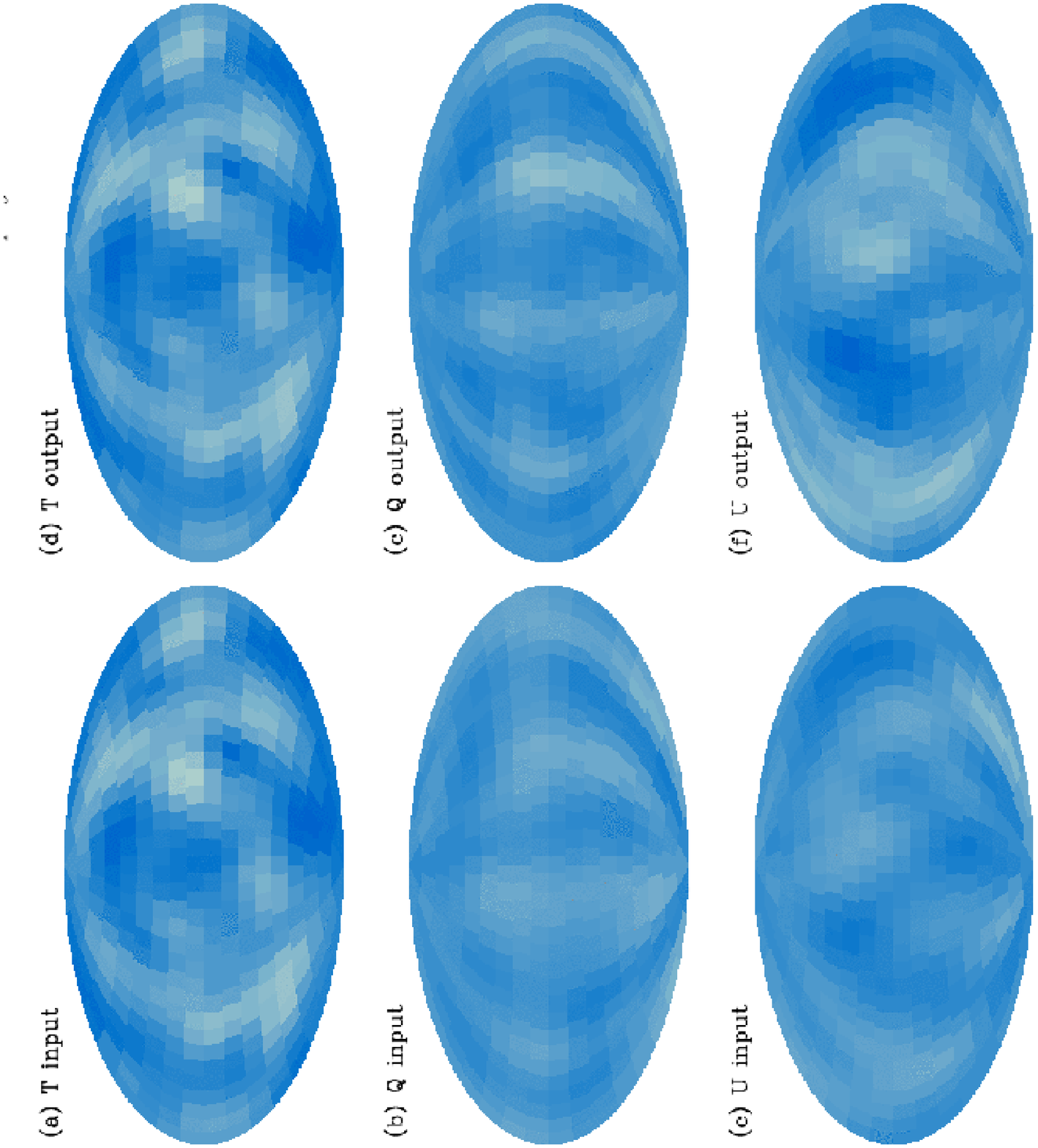}

\caption
{The maps from Figure \ref{figure7} smoothed with a Gaussian beam of
FWHM $20^\circ$ and pixelised into maps with $10^\circ \times
10^\circ$ pixels. The \planck-like polarization maps (e) and (f) are
dominated by signal at low multipoles (see Figures \ref{figure6}c and
\ref{figure6}d), but noise from residual striping errors is also
visible at this resolution.}

\label{figure8}

\end{figure*}

Figure \ref{figure6} shows the power spectra for the realisation of
Figure \ref{figure5} but now including destriping and white noise as
described above. The maps corresponding to these power spectra are
shown in the right hand panels of Figure \ref{figure7}. The white
noise level for these simulations is so low that it cannot be seen in
the $T$ and $X$ power spectra. However, the $E$ and $B$ mode power
spectra are dominated by white noise at $\ell \simgt 20$. The $E$ mode
power spectrum is signal dominated at $\ell \simlt 10$ and at these
low multipoles the dominant source of noise comes from destriping
errors. Since the white noise for each detector is assumed to be the
same, the noise contributions to the $E$ and $B$ power spectra are
almost identical. Comparing the right hand panels of Figure
\ref{figure7} with the input maps, one can see that the $Q$ and $U$
maps are dominated by white noise, but some large-scale CMB features 
can be discerned in these noisy maps.

Figure \ref{figure8} shows the maps of Figure \ref{figure7} smoothed
with a Gaussian of width $\theta_s = 8.5^\circ$ and repixelised onto
pixels of width $\Delta \theta_c = 10^\circ$. The large scale CMB
features in the output $Q$ and $U$ maps are now clearly visible and
the dominant source of errors are stripes aligned (almost) along great
circles in the sky. If \plancks works as expected ({\it i.e.} no
unexpected and `uncorrectable' sources of low frequency noise) the
polarization maps at this frequency should be signal dominated on
large scales.

\begin{figure*}

\vskip 5.3 truein

\includegraphics{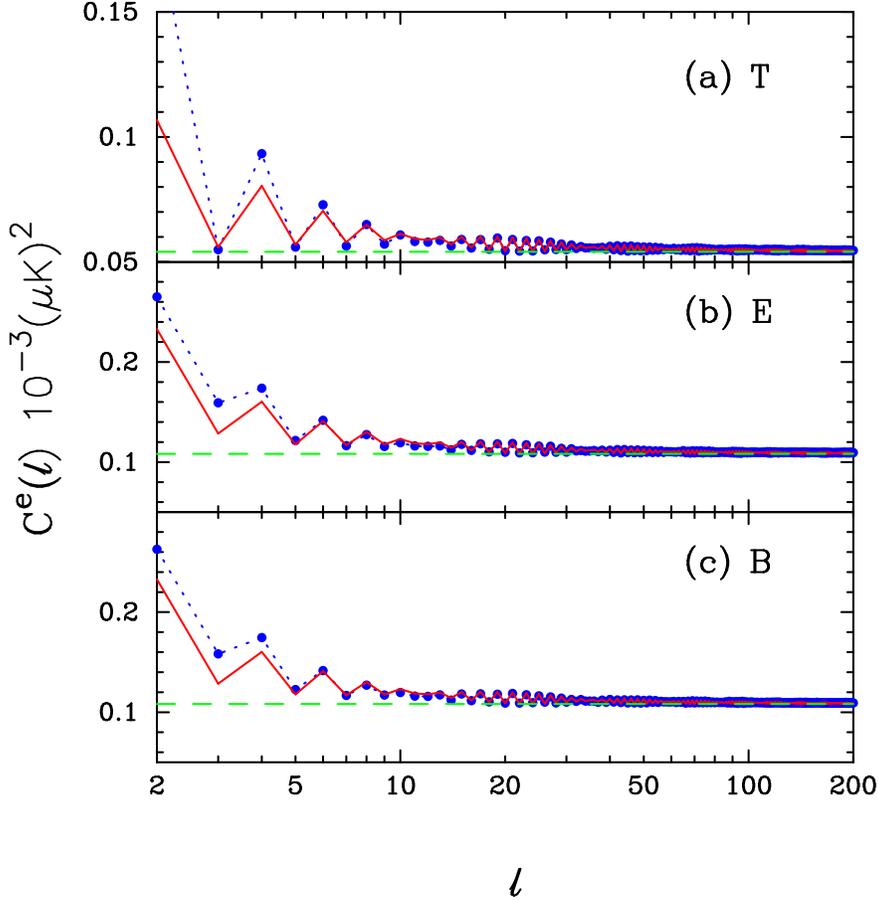}

\caption
{The (blue) circles show the mean error power spectra for $4000$
simulations of the type shown in right hand panels of Figure
\ref{figure7}.The (green) dashed line shows the white noise level to
which these power spectra  converge at high multipoles. The (red) solid
lines shows sum of the white noise levels and the analytic
predictions of equations (\ref{RT12a})- (\ref{RT12c}) for a ring-torus
geometry with boresight angle $86.15^\circ$.}

\label{figure9}

\end{figure*}

Figure \ref{figure9} shows the error power spectra determined from
$4000$ simulations with parameters as described above (plotted as the
blue circles). The (green) dashed lines in the Figure show the white
noise level expected for each power spectrum. The (red) lines show the
ring-torus model of equations (\ref{RT12a}) - (\ref{RT12c}) scaled to
provide a good fit to the numerical results. (A boresight angle of
$86.15^\circ$, appropriate for the polarized $143$GHz \plancks
detectors, has been used for these predictions.) The ring-torus model
deviates slightly from the numerical results at $\ell \simlt 5$, but
provides a highly accurate match to the error power spectra at higher
multipoles. It is therefore feasible to use the error model to fit and
subtract the noise biases to the power spectrum estimates for the $E$
and $B$ modes at $\ell \simgt 5$. It is also worth pointing out that
the error power spectra will increase the variances of the power
spectrum estimates approximately as
\begin{equation}
   \langle (\Delta C_\ell)^2 \rangle \approx { 2 \over ( 2 \ell + 1)} (C^{\rm 
CMB}_\ell + C^e_\ell)^2,
\end{equation}
and will introduce coupling between coefficient of different $\ell$
(see below).

\begin{figure*}

\vskip 8.0 truein

\includegraphics{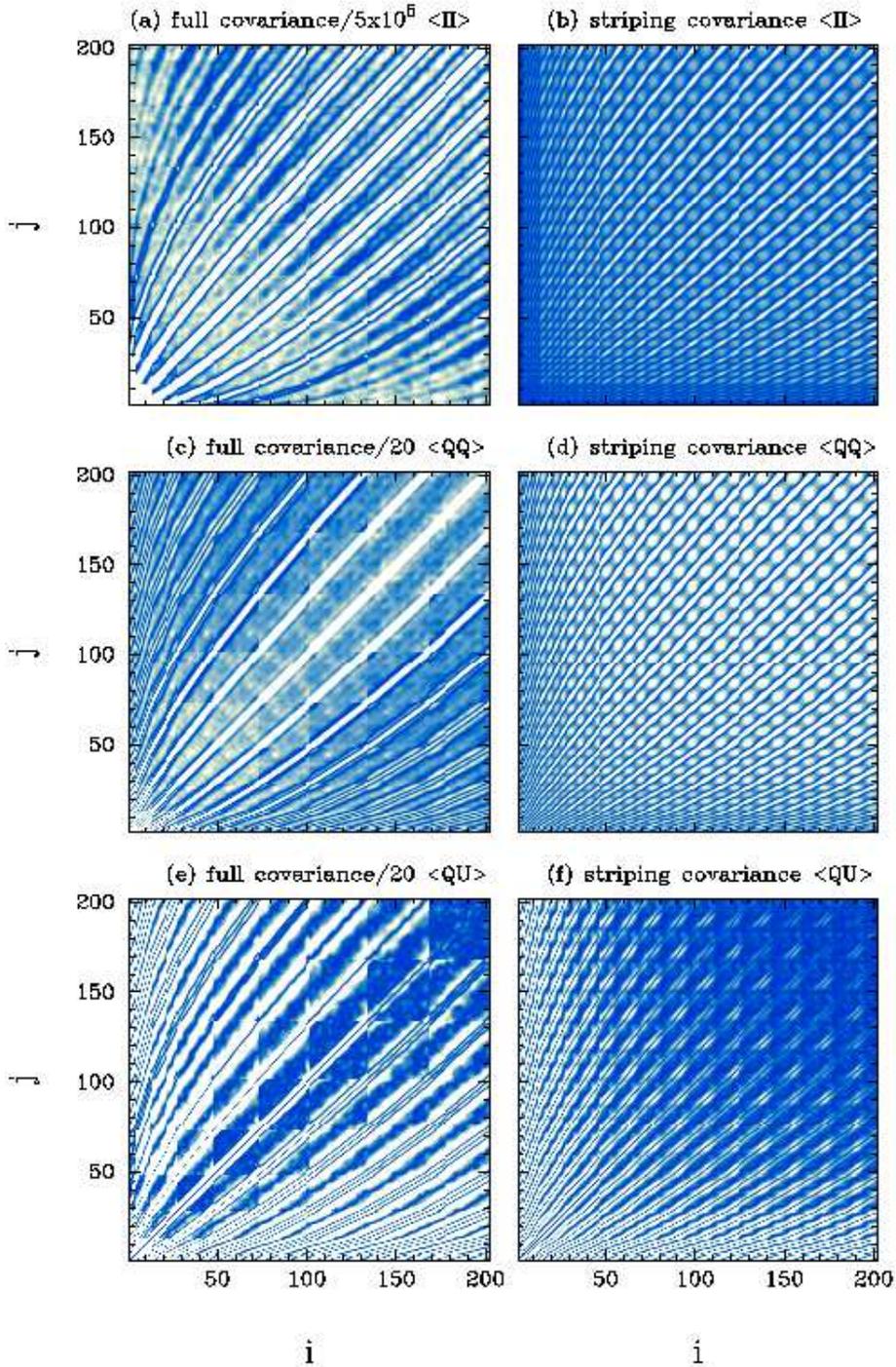}

\caption
{The panels to the left show one quadrant of the full pixel-pixel
noise covariance matrices ({\it i.e.} signal+destriping errors+white
noise) estimated from $4000$ simulations of degraded resolution
destriped maps such as those shown in Figures \ref{figure8}d --
\ref{figure8}f. The panels to the right show the contribution to these
covariance matrices arising from destriping errors alone. These can be
compared to the ring-torus model plotted in Figure \ref{figure4}.
Note that the full covariance matrices have been divided by scaling
factors to render them visible: the $\langle II \rangle$ covariance
plotted in (a) has been divided by a factor of $5 \times 10^5$; the
$\langle QQ \rangle$ and $\langle QU \rangle$ covariances in (c) and
(e) have each been divided by a factor of $20$.}
\label{figure10}

\end{figure*}

Pixel-pixel covariance matrices averaged over $4000$ degraded
resolution maps, such as those plotted in Figure \ref{figure4}, are shown
in Figure \ref{figure10}. The left hand panels show the full
pixel-pixel covariance matrices, which include signal, destriping
errors and white noise. The panels to the right show the contribution
to the pixel noise arising from destriping errors alone. These can be
compared to the equivalent figures for the ring torus geometry shown
in Figure \ref{figure4}. Apart from differences in the amplitudes
arising from differences between the dispersion of the ring offsets,
the additional structure seen in Figure \ref{figure10} is caused by 
the slow precession in the scanning strategy.

\begin{figure*}

\vskip 3.0 truein

\includegraphics{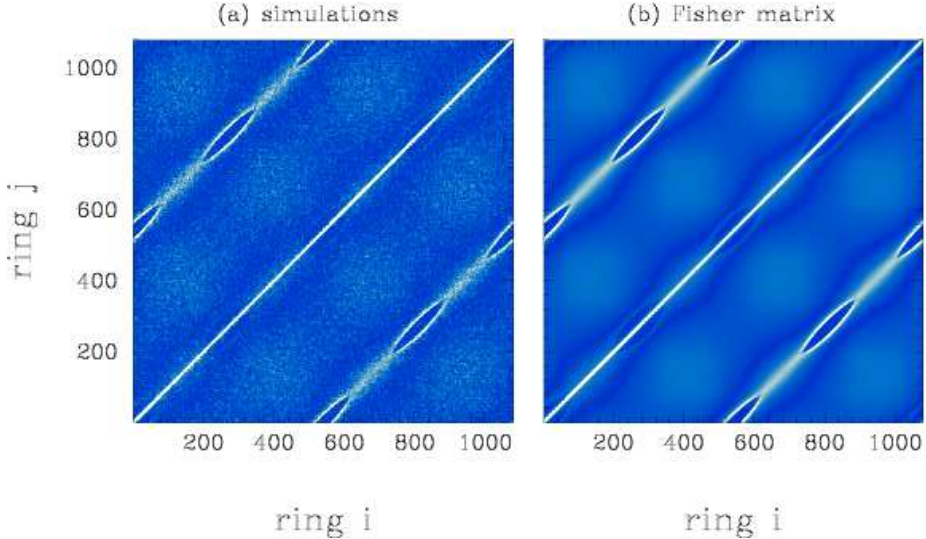}

\caption
{Figure \ref{figure11}a shows the covariance matrix of the 
destriped ring offsets for a single bolometer determined from
4000 simulations. Figure \ref{figure11}b shows the theoretical covariance
matrix computed from equation (\ref{Sim3}).  }

\label{figure11}

\end{figure*}

The left hand panel of Figure \ref{figure11} shows the coviariance matrix
for the ring offsets $\langle a_ia_j \rangle$ for one detector determined
by averaging over $4000$ simulations. The ring offset covariance matrices
are dominantly diagonal, as assumed in the ring torus model of Section 3, 
but there are some low amplitude correlations particularly with the rings
separated by $\Delta \phi \approx \pi$ in ecliptic longtitude. The panel
to the right shows the covariance matrix given by the inverse of the Fisher
matrix computed from the $\tilde \chi^2$ of equation (\ref{Sim1}):
\begin{equation}
\left. \begin{array}{ccc}
F_{k k}   = & {\partial^2 \tilde \chi^2 \over \partial a_k^2} =& \sum_p \sum_{ik\subset p} {1 \over
\sigma^2 n_p}  + 2 \lambda,  \\
F_{k k^\prime} = &{\partial^2 \tilde \chi^2 \over \partial a_k \partial a_k^\prime}
= & \sum_p \sum_{ijkk^\prime\subset p} {-1 \over
\sigma^2 n_p}  + 2 \lambda, \end{array} \right \} \label{Sim3} 
\end{equation}
which depends only on the scanning pattern and the TOD variance
$\sigma^2$. The theoretical covariance matrix (\ref{Sim3}) provides an
excellent match to the empirical covariance matrix derived from
destriping.

Using $1/8$th of the number of rings for \plancks it is feasible, as
shown in this Section, to analyse noise power-spectra and estimate the
pixel-pixel noise covariance matrices of low resolution maps by brute
force simulations of the TOD. At full \plancks resolution this will
become computationally challenging, if not impossible. For example,
suppose we wish to analyse low resolution \plancks maps consisting of
$10^4$ pixels, one would probably require more than $10^5$ simulations
of the TOD at each frequency to accurately characterise the pixel
noise covariance matrices. However, Figure \ref{figure11} suggests
that full simulations of the TOD are not required. Instead, we can
decompose the noise into an `uncorrelated' component described by the
$3 \times 3$ matrices of equation (\ref{TOD10}) and a correlated
component quantifying the destriping errors given by the Fisher
matrix of (\ref{Sim3}). We will focus on modelling the destriping
errors in the remainder of this Section.

\begin{figure*}

\vskip 8.0 truein

\includegraphics{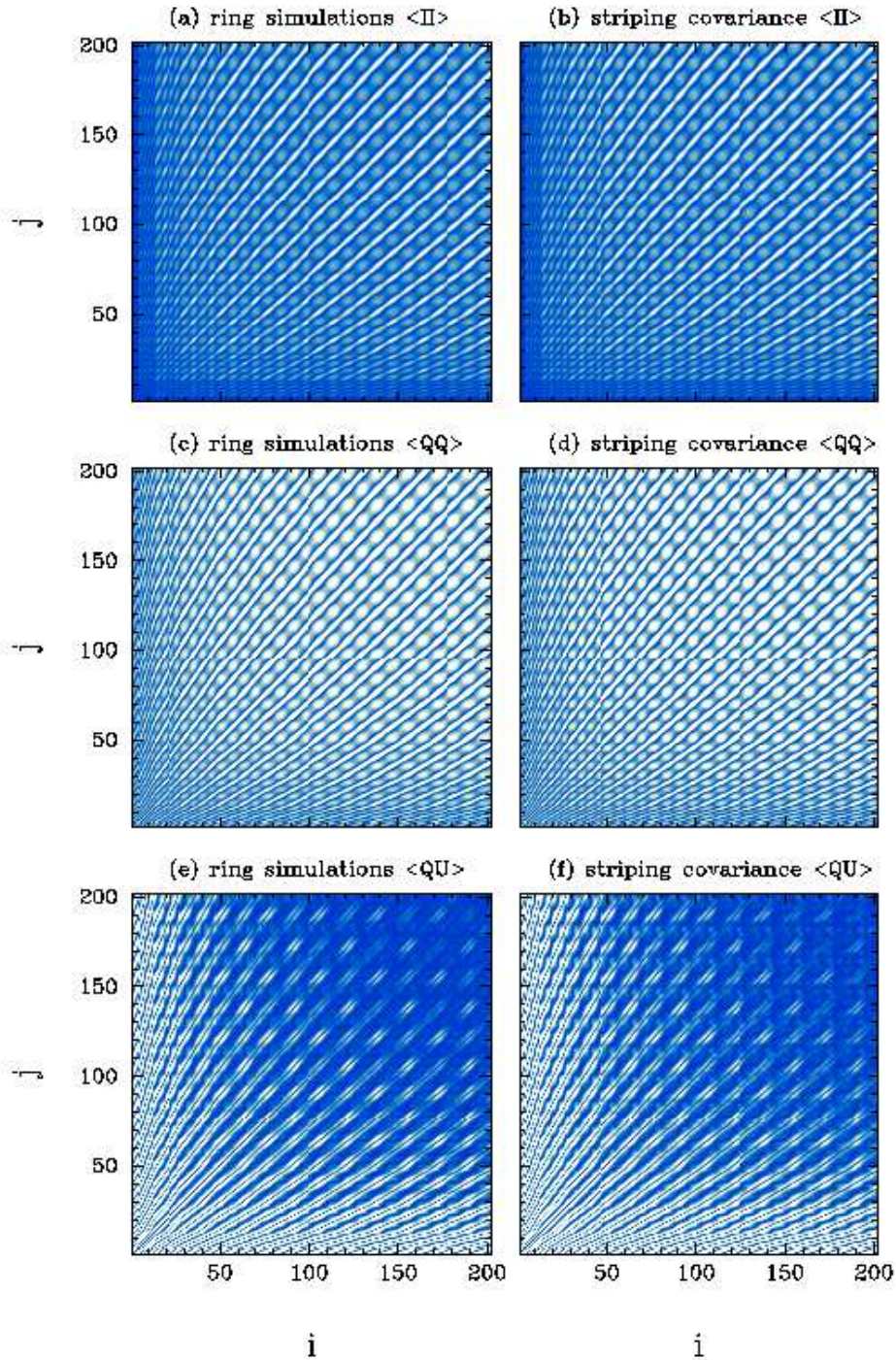}

\caption
{The right hand panels are identical to the right hand
panels of Figure \ref{figure11} and show the
pixel-pixel covariance matrices arising from destriping errors
alone. The left hand panels show the pixel covariance matrices 
from $10,000$ simulations generated from Gaussian realisations
of ring offsets with the theoretical covariance matrices of
equation (\ref{Sim3}). }

\label{figure12}

\end{figure*}

Given the Fisher matrix (\ref{Sim3}) it is straightforward, and very
fast, to generate simulations of correlated baseline coefficients and
to compute $I$, $Q$ and $U$ maps and error power spectra. These maps
can then be degraded to a lower resolution after smoothing by a
Gaussian (\ref{Beam1}) or a more complex function (for example, if one
wants to minimise contamination from the Galaxy). An example is shown
in Figure \ref{figure12}. In this Figure, the left hand panels are
identical to the right hand panels of Figure \ref{figure10} show the
destriping contributions to the pixel covariances derived from full
simulations of the TOD. The panels to the right show the destriping 
covariances from ring simulations assuming that the ring coefficients
are Gaussian distributed with a covariance matrix $F_{k k^\prime}^{-1}$.
The results are essentially identical to those from the simulations of the
TODs. 

Monte-Carlo simulations of the baseline offsets is therefore provide
an entirely feasible approach to analysing the complex striping noise
for \planck. The map comparisons in A06 show that for realistic `$1/f$'
detector noise it is possible to get close to the `irreducible'
destriping errors using a small number of baseline offsets (as
mentioned in the introduction, one offset per ring comes close to the
`optimal' map solution) and so a manageable number of baseline offsets
can be used in the Monte-Carlo simulations. The example shown in Figure
\ref{figure6} shows that the $E$-mode power spectrum for the $143$GHz
\plancks channel should be signal dominated at multipoles $\ell \simlt
10$ and at $\ell \simlt 20$ destriping errors will be the dominant
source of noise. It is therefore important to model the destriping
noise from \planck.  The amplitude of $B$-mode spectrum is unknown
(for a recent review see Efstathiou and Chongchitnan, 2006) and may
well be lower than the \plancks detection threshold, corresponding to a
tensor-scalar ratio $r \sim 0.1$ (SPP05) An accurate analysis of
destriping errors will therefore be essential for the analysis of
primordial $B$-mode anistropies (particularly for the possible
detection of a `reionisation bump' at $\ell \simlt$, which cannot
be observed in ground based experiments).

Finally, we mention that if the CMB fluctuations are an isotropic
Gaussian random field, then we expect the harmonic coefficients to
satistfy
\begin{equation}
\langle a_{\ell m} a^*_{\ell m^\prime} \rangle = C_\ell \delta_{\ell \ell^\prime} \delta_{m m^\prime}. \label{Sim4}
\end{equation}
The simulations described above show that the destriping errors are
well approximated by a Gaussian random field and, of course, Gaussianity
is assumed in generating simulations from the covariance matrices of the
ring offsets. However, the striping errors, although Gaussian, are
evidently anisotropic since they are highly correlated along the scan
direction.  Failure to model the destriping errors could therefore
have a significant effect on tests for non-Gaussianity in the \plancks
data. The striping errors will also introduce correlations in the harmonic
coefficients. For example, for the temperature anisotropies, E05 used the
ring torus model to show that the striping errors lead to correlations:
\begin{equation}
\langle a^e_{\ell m} a^{*e}_{\ell^\prime m^\prime} \rangle = 
{\pi \over 4}\sigma^2 \sin^2 \theta_b \Delta \alpha \delta_{m m^\prime}
\sum_{m_1 m_2} \hat P_{\ell m_1} \hat P_{\ell^\prime m_2} I(m_1) I(m_2) 
d^\ell_{m_1 m} d^{\ell^\prime}_{m2 m^\prime} (\pi /2), \label{Sim5}
\end{equation}
violating the statistical isotropy of equation (\ref{Sim4}). The ring
torus model of Section 3 can be used to show that destriping errors
introduce similar correlations in the polarization harmonic
coefficients. In assessing non-Gaussianity, using either map based
statistics or statistics based on harmonic coefficients, it will be
important to model the effects of destriping errors. This can be done
at full \plancks resolution by generating large numbers of simulations
using ring covariance matrices, rather than simulations of the TODs.

\section{Conclusions}

This paper has presented an analysis of destriping errors in
temperature and polarization maps for a \planck-like experiment,
generalising the results of E05 to polarization. For a simple detector
geometry and ring torus scanning pattern, we have shown that it is
possible to compute analytically the effects of striping errors on
both the power spectra and pixel-pixel covariances.

In Section 4 we have compared the ring torus model against simulations
adopting the focal plane geometry of the \plancks polarized $143$GHz
detectors and a slowly precessing scan strategy. The ring torus model
provides a very accurate description of the noise power spectra at
$\ell \simgt 5$.  The white noise in the detector TODs in the
numerical simulations was chosen to match the expected noise levels of
the \plancks detectors in $30^\prime \times 30^\prime$ pixels. Thus at
low multipoles, the simulations should give a good indication of the
expected performance of \planck. The simulations show that \plancks
polarization maps should be signal dominated at large scales. The
$E$-mode power spectrum should be signal dominated at $\ell \simlt 10$
and destriping errors will be the dominant source of noise in the $E$-
and $B$-mode power spectra at $\ell \simlt 20$. Destriping errors will
therefore need to be modelled accurately in the analysis of the
polarization power spectra from \planck.

As discussed in E06, since the noise is expected to be white at high
multipoles, near-optimal estimates of the power spectra and their
covariances can be obstained using a hybrid power spectrum estimator.
Provided one has a model for the pixel covariance matrices,  the low
multipoles can be determined by applying a maximum likelihood
estimator to low resolution maps. Section 4 shows that a model for the
pixel covariances for maps of arbitrary resolution can be constructed
accurately using fast simulations based on the statistical properties
of the destriping baseline offsets,  rather than simulations of the full
TOD. Such fast simulations will be useful for other purposes, for
example, in assessing the signficance of tests for non-Gaussianity in
the \plancks maps.

\vskip 0.1 truein

\noindent
{\bf Acknowledgements:} I thank members of the Cambridge \plancks
Analysis Centre, especially Anthony Challinor and Mark Ashdown, for
helpful discussions.


\begin{thebibliography}{}

\bibitem[\protect\citename{Ash}2006]{Ash06}
Ashdown M.A.J., \etal,  2006, submitted to A\&A. astro-ph/0606348.


\bibitem[\protect\citename{BFJS}2001]{BFJS01}
Borrill J., Ferreira P. G., Jaffe A. H., Stompor R., 2001, In {\it
`Mining the Sky'}, Proceedings of the MPA/ESO/MPE Workshop, Edited by 
A. J. Banday, S. Zaroubi, and M.Bartelmann. Springer-Verlag, Heidelberg,  p403.

\bibitem[\protect\citename{BS93}1993]{BS93}
Brink D.M., Satchler G.R,. 1993, {\it Angular Momentum}, third edition, Oxford University
Press, Oxford.

\bibitem[\protect\citename{BMMDMBM97}1997]{BMMDMBM97}
Burigana C., Malaspina M., Mandolesi N., Danese L., Maino D., Bersanelli M.,
Maltoni M., 1997, Internal Report ITESRE. astro-ph/9906360.




\bibitem[\protect\citename{CC05}2005]{CC05}
Challinor A.D.,  Chon G., 2005,  MNRAS, 360, 509. 

\bibitem[\protect\citename{D98}1998]{D98}
Delabrouille, J.,  1998,  A\&A Suppl. Ser., 127, 555.

\bibitem[\protect\citename{DK01}2002]{DK02}
Delabrouille, J., Kaplan J.,   2002,  Proceedings of the Pol2001
Astrophysical Polarised Backgrounds' conference, 
Bologna, AIP Conference Proceedings, 609, 135.


\bibitem[\protect\citename{DKP01}2001]{DKP01}
Dor\'e O., Teyssier R., Bouchet F.R., Vibert D., Prunet S., 2001,
A\&A, 374, 358.

\bibitem[\protect\citename{E04}2004]{E04}
Efstathiou G., 2004,  MNRAS, 439, 603.

\bibitem[\protect\citename{E05}2005]{E05}
Efstathiou G., 2005,  MNRAS, 356, 1549.

\bibitem[\protect\citename{E06}2006]{E06}
Efstathiou G., 2006,  MNRAS, 370, 343.

\bibitem[\protect\citename{E06}2006]{E06}
Efstathiou G.,  Chongchitnan S. 2006,   Prog. Theor. Phys. Suppl., 163, 204.

\bibitem[\protect\citename{Hetal2}2002]{Hetal02}
Hivon E., G\'orski K.M., Netterfield C.B., Crill B.P., Prunet S., Hansen F.,
2002, ApJ, 567, 2.

\bibitem[\protect\citename{KKS}1997]{KKS97}
Kamionkowski M., Kosowsky A.,  Stebbins A., 1997, PRD, 55, 7368.


\bibitem[\protect\citename{KKPMB03}2003]{KKPMB03}
Keih\"anen E., Kurki-Suonio H., Poutanen T., Maino, D., Burigana C., 
2004,   A\&A, 438, 287.


\bibitem[\protect\citename{KKP05}2005]{KKP05}
Keih\"anen E., Kurki-Suonio H., Poutanen T., 
2005,  MNRAS, 360, 390.




\bibitem[\protect\citename{Letal06}2006]{Letal06}
Larson D.L., Eriksen H.K., Wandelt B.D., G\'orski K.M., Huey G., Jewell J.B.,
O'Dywer I.J., 2006, ApJ, in press. astro-ph/0608007.


\bibitem[\protect\citename{LC06}2006]{LC06}
Lewis A., Challinor, A., 2006, Phys. Rep.,  429,  1.


\bibitem[\protect\citename{Metal99}1999]{Metal99}
Maino D., {\it et al.}, 1999,  A\&A Suppl. Ser., 140, 383.

\bibitem[\protect\citename{NdeGGV01}1997]{NdeGGV01}
Natoli P., de Gasperis G., Gheller C., Vittorio N., 2001, A\&A, 372, 
346.

\bibitem[\protect\citename{PO5}2005]{P05}
The \plancks Consortia,  2005,  {\it `The Scientific Programme of \planck'},
ESA-SCI(2005)1, European Space Agency. astro-ph/0604069. (SPP05)

\bibitem[\protect\citename{RKACDK00}2000]{RKACD00}
Revenu B., Kim A., Ansari R., Couchot F., Delabrouille J., Kaplan J.,
2000,  A\&A Suppl. Ser., 142, 499.

\bibitem[\protect\citename{SS04}2004]{SS04}
Slosar A., Seljak, U., 2004, PRD, 70, 3002.

\bibitem[\protect\citename{Setal06}2006]{Setal6}
Spergel D., \etal, 2006, submitted to ApJ. astro-ph/0603449.



\bibitem[\protect\citename{T97a}1997]{T97a}
Tegmark M., 1997a, ApJL, 480, L87.

\bibitem[\protect\citename{T97b}1997]{T97b}
Tegmark M., 1997b, PRD, 56, 4514.

\bibitem[\protect\citename{T97c}1997]{T97c}
Tegmark M., 1997c, PRD, 55, 5895.

\bibitem[\protect\citename{TdO97}2001]{TdO01}
Tegmark M., de Oliveira-Costa A.,  2001, PRD, 64, 063001. 

\bibitem[\protect\citename{VMK88}1988]{VMK88} Varshalovich, D.A.,
Moskalev A.N., Khersonskii V.K., 1988, {\it Quantum Theory of Angular
Momentum}, World Scientific, Singapore.


\bibitem[\protect\citename{WHB96}1996]{WHB96}
Wright E.L., Hinshaw G., Bennett C.L.,  1996, ApJ, 458, L53.


\end{thebibliography}
\end{document}